\documentclass[a4paper]{article}
\usepackage{lmodern}
\usepackage{amssymb,amsmath}
\usepackage{ifxetex,ifluatex}
\usepackage{fixltx2e} % provides \textsubscript
\ifnum 0\ifxetex 1\fi\ifluatex 1\fi=0 % if pdftex
  \usepackage[T1]{fontenc}
  \usepackage[utf8]{inputenc}
\else % if luatex or xelatex
  \ifxetex
    \usepackage{mathspec}
  \else
    \usepackage{fontspec}
  \fi
  \defaultfontfeatures{Ligatures=TeX,Scale=MatchLowercase}
\fi
\IfFileExists{upquote.sty}{\usepackage{upquote}}{}
\IfFileExists{microtype.sty}{%
\usepackage{microtype}
\UseMicrotypeSet[protrusion]{basicmath} % disable protrusion for tt fonts
}{}
\usepackage[margin=1in]{geometry}
\usepackage{hyperref}
\hypersetup{unicode=true,
            pdfborder={0 0 0},
            breaklinks=true}
\usepackage{longtable,booktabs}
\usepackage{graphicx,grffile}
\makeatletter
\def\maxwidth{\ifdim\Gin@nat@width>\linewidth\linewidth\else\Gin@nat@width\fi}
\def\maxheight{\ifdim\Gin@nat@height>\textheight\textheight\else\Gin@nat@height\fi}
\makeatother
\setkeys{Gin}{width=\maxwidth,height=\maxheight,keepaspectratio}
\IfFileExists{parskip.sty}{%
\usepackage{parskip}
}{% else
\setlength{\parindent}{0pt}
\setlength{\parskip}{6pt plus 2pt minus 1pt}
}
\providecommand{\tightlist}{%
  \setlength{\itemsep}{0pt}\setlength{\parskip}{0pt}}
\ifx\paragraph\undefined\else
\let\oldparagraph\paragraph
\renewcommand{\paragraph}[1]{\oldparagraph{#1}\mbox{}}
\fi
\ifx\subparagraph\undefined\else
\let\oldsubparagraph\subparagraph
\renewcommand{\subparagraph}[1]{\oldsubparagraph{#1}\mbox{}}
\fi

\let\rmarkdownfootnote\footnote%
\def\footnote{\protect\rmarkdownfootnote}

\usepackage{titling}

  \title{}
  \pretitle{\vspace{\droptitle}}
  \posttitle{}
  \author{}
  \preauthor{}\postauthor{}
  \date{}
  \predate{}\postdate{}

\usepackage{amsmath}
\usepackage{amsfonts,amssymb,textcomp,gensymb}
\usepackage{bbm}

\newcommand{\Rset}{\mathbb{R}}
\newcommand{\Xset}{\mathbb{X}}

\newcommand{\x}{\mathbf{x}}

\usepackage{scrpage2}
\ifoot[]{}
\ofoot[\pagemark]{\pagemark}

\usepackage{lscape}
\newcommand{\landscapebegin}{\begin{landscape}}
\newcommand{\landscapeend}{\end{landscape}}

\begin{document}

\section{Using numerical plant models and phenotypic correlation space
to design achievable
ideotypes.}\label{using-numerical-plant-models-and-phenotypic-correlation-space-to-design-achievable-ideotypes.}

Victor Picheny (1)*, Pierre Casadebaig (2)*, Ronan Trépos (1), Robert
Faivre (1), David Da Silva (3), Patrick Vincourt (4), Evelyne Costes (3)

\begin{enumerate}
\def\labelenumi{(\arabic{enumi})}
\tightlist
\item
  INRA, UR875 MIAT, 31326 Castanet-Tolosan, France
\item
  INRA, UMR1248 AGIR, 31326 Castanet-Tolosan, France
\item
  INRA, UMR1334 AGAP CIRAD-INRA-Montpellier SupAgro, 34060 Montpellier,
  France
\item
  INRA, UMR441 LIPM, 31326 Castanet-Tolosan, France
\end{enumerate}

(*) The first two authors contributed equally to this work.

\section[Abstract]{\texorpdfstring{Abstract\footnote{This is the
  post-print version of the manuscript published in \emph{Plant, Cell,
  and Environment}
  (\href{http://dx.doi.org/10.1111/pce.13001}{10.1111/pce.13001})}}{Abstract}}\label{abstract}

\begin{quote}
Numerical plant models can predict the outcome of plant traits
modifications resulting from genetic variations, on plant performance,
by simulating physiological processes and their interaction with the
environment. Optimization methods complement those models to design
ideotypes, i.e.~ideal values of a set of plant traits resulting in
optimal adaptation for given combinations of environment and management,
mainly through the maximization of a performance criteria (e.g.~yield,
light interception). As use of simulation models gains momentum in plant
breeding, numerical experiments must be carefully engineered to provide
accurate and attainable results, rooting them in biological reality.\\
Here, we propose a multi-objective optimization formulation that
includes a metric of performance, returned by the numerical model, and a
metric of feasibility, accounting for correlations between traits based
on field observations. We applied this approach to two contrasting
models: a process-based crop model of sunflower and a
functional-structural plant model of apple trees. In both cases, the
method successfully characterized key plant traits and identified a
continuum of optimal solutions, ranging from the most feasible to the
most efficient.\\
The present study thus provides successful proof of concept for this
enhanced modeling approach, which identified paths for desirable trait
modification, including direction and intensity.
\end{quote}

\newpage

\section{Introduction}\label{introduction}

\subsection{Using simulation models to optimize
phenotype}\label{using-simulation-models-to-optimize-phenotype}

Global demand for agricultural products to supply food, feed, and fuel
is rapidly increasing (Edgerton, 2009). One option to meet this growing
agriculture need is to continue improving plant productivity per unit of
cultivated land area. However, after decades of increase, many major
crops have recently shown slower rates of yield improvement, stagnation,
or even loss of productivity (Ray et al., 2012). This situation likely
results from the negative effects of global climate change and societal
prejudice that perceives environmental costs and limits agricultural
resources, for example, public policies to reduce the use of chemicals
in disease management programs (Sutton, 1996). Overcoming these
challenges will require that breeders continue the genetic improvement
of major crops accounting for changing agricultural practices towards
sustainable production systems (Vanloqueren and Baret, 2009).

In general, selecting plant traits associated with improvements in crop
yield is difficult. Improvements at the organ or plant level must scale
to the field level and these increases must not be gained at the expense
of other traits that may ultimately offset yield advances (Sinclair et
al., 2004). In addition, the specific environment where the crop is
grown has a significant effect on yield. These genotype \(\times\)
environment (G \(\times\) E) interactions require considerable
experimental effort to identify the persistent traits that contribute to
yield increases across many environments. The effort required is
redoubled in pluri-annual crops where data, collected over changing
environments, are difficult to interpret for complex phenotypes or at a
fine scale. Computer-based modeling approaches have recently emerged as
a method to save time, labor, and resources and to infer traits value
beyond field experiments (Casadebaig, Zheng, et al., 2016 ; David Da
Silva et al., 2014; Martre, He, et al., 2015).

This computer-based strategy has been developed over the past two
decades using a number of simulation models. These software models
derive from mathematical equations that represent the biological
processes that influence plant growth and development as a function of
time, environment (climate, soil, and management), and
genotype-dependent parameters. Those parameters are expected to be more
heritable than complex traits, less prone to G x E interactions and to
have a less complex genetic architecture (Heslot et al., 2014). These
parameters also represent a range of functional traits, i.e any
morphological, physiological, phenological or behavioral feature that is
measurable at the individual level (Violle et al., 2007). In this case,
simulation is used as a tool to predict trait \(\times\) trait and trait
\(\times\) environment interactions when scaling from the individual
plant to the field level.

Depending on their structure, simulation models can be referred to as
\emph{process-based crop models} or \emph{functional-structural plant
models}. Process-based models are defined at the plot scale, often
without an explicit representation of individual plants and focus on
predicting crop performance, mainly harvestable organ yield or quality.
Differences in the physiological framework (e.g radiation-based or
water-based biomass production) lead to distinct families of
process-based model that have been used worldwide (Brisson et al., 2003;
Keating et al., 2003; Ritchie and Otter, 1985; Stockle et al., 2003).
Such models are used to explore the G \(\times\) E landscape and assist
breeding programs by taking advantage of genetic and environmental
resources (e.g. Chapman et al., 2003; Hammer et al., 2006; Jeuffroy et
al., 2014). By contrast, functional-structural plant models particularly
represent the plant structure (i.e.~its topology and geometry) linked
with the functions that allow the plant to interact with its
environment, i.e.~light interception, photosynthesis, carbon allocation,
etc. (DeJong et al., 2011). Depending on the objectives, these models
focus on specific plant structure such as trunk and wood (Nikinmaa et
al., 2003) vegetative development (Costes et al., 2008; Fournier and
Andrieu, 1999; Lopez et al., 2008), or fruit development (Allen et al.,
2005).

The concept of \emph{ideotype} describes the idealized realization of a
plant phenotype, ``a biological model which is expected to perform or
behave in a predictable manner within a defined environment'' (Donald,
1968). This definition provides the breeder with a guide for
characterization of cultivar by identifying and selecting for particular
features in progenies. Model-assisted phenotyping and ideotype design is
a research domain with recent developments (Martre, Bénédicte
Quilot-Turion, et al., 2015). These advances formulates ideotype design
as an optimization of model inputs related to plant traits (David Da
Silva et al., 2014; Ding et al., 2016; Paleari et al., 2015;
Quilot-Turion et al., 2012; Semenov et al., 2014; Semenov and
Stratonovitch, 2013), allelic combinations (e.g. Letort et al., 2008;
Quilot-Turion et al., 2016), or management options (Grechi et al., 2012;
Wu et al., 2012). These approaches have been used for different goals,
including understanding crop adaptation to climate change (Paleari et
al., 2015; Semenov et al., 2014; Semenov and Stratonovitch, 2013),
sustainable production (Quilot-Turion et al., 2012) or integrated pest
management (Grechi et al., 2012).

\subsection{Research objectives and
challenges}\label{research-objectives-and-challenges}

Despite the observation of trait correlation, i.e.~traits covary among
individuals within a population, current model-based ideotype designs do
not formally consider these correlations. This failure to account for
trait correlations can lead to numerical experiments that are not well
rooted in biological reality and ultimately fail to provide meaningful
data to the breeder. These correlations are observed at the phenotypic
level and result from differences in the variances and covariances at
the genetic (G) and environmental (E) level (and their interactions,
GxE), or between consecutive years or branching orders in perennial
plants (Segura et al., 2008). Multi-environmental trials allow
separation of the variance and covariance components for G, E and GxE.
However, this process is generally based on the assumption that
covariance structure for E does not depend on the genotype, and vice
versa.

Here, we propose a notion of \emph{feasibility}, that can be viewed as
the probability that the breeder will be able to find progenies having
the expected ideotype from a cross between given parents, based on
meiosis. The distribution of a trait reflects its dispersion and the
possibility of its selection. Thus, many individuals will have values
close to the mean value of the progeny while some individuals will have
a phenotype more extreme than the parents (heterotic or transgressive
traits). In this perspective, some trait associations may be unlikely
due to negative correlations between the traits, possibly from
physiological antagonisms. Other traits may be difficult to dissociate
due to close proximity of their respective key genes (linkage), which
can result in a lack of genetic recombinants. In addition, several genes
may interact and have a combined effect on a given trait (epistasis) or
one gene may simultaneously affect several traits (pleiotropy). Thus,
the architecture of variances and correlations between traits reveals
the feasibility of jointly selecting traits to bolster genetic gain.

In our approach, we define a feasibility criterion, based on the
observed joint distribution of the traits used as model parameters, as a
possible means of improving the validity of ideotype design. Introducing
real data is a means to control the optimization process and can be used
to improve predictions similar to data assimilation for dynamic models.
From an operational point of view, we focus on \emph{empirical}
correlations observed at the plant level (in agronomic plant
populations); the aforementioned distinctions between genetic and
environmental variances are not considered in this work. Indeed, due to
the sample size, these empirical, phenotypic correlations were expected
to be more robust than inferred genetic correlations.

The present work evaluated an apple tree orchard and a sunflower crop to
design realistic and efficient plant ideotypes. The study relied on two
numerical models: a functional-structural plant model for the apple tree
(MAppleT, Costes et al., 2008) and a process-based model for the
sunflower crop (SUNFLO, Casadebaig et al., 2011; Lecoeur et al., 2011).
These examples targeted different steps in the breeding process and
considered different sources for genetic variability:(1) upstream
sources, with a segregating population of apple tree, and (2) downstream
sources, with a collection of sunflower cultivars. In both cases, we
used a unified multi-objective optimization formulation to solve the
ideotyping problem. We used a metric of performance returned by the
numerical model, and the introduced metric of feasibility based on
observations. The main objective of our approach was to improve the
realism of model-based ideotype design, and no modification of the
simulation models was required: the approach we propose is therefore
suitable for different types of simulation models (process-based,
functional structural, etc.) or input data.

\section{Materials and methods}\label{materials-and-methods}

\subsection{Apple tree orchards}\label{apple-tree-orchards}

Breeding programs for apple have primarily focused on major traits, such
as disease resistance and fruit quality, despite the desirability of
additional traits, such as bearing regularity or an optimized fruit
distribution in the canopy (Laurens et al., 2000; Lespinasse et al.,
1992 ). More recently, new selection criteria have been proposed that
allow adaptation to climate change. Such traits include tree
architecture (leaf area, branching), phenology (flowering, vegetative
shoot, and fruit maturation), and the ability to tolerate periods of
water deprivation during the growing season.

The architecture of a tree determines the three-dimensional foliage
distribution and consequently the efficiency of light interception. It
thereby impacts water transport and transpiration as well as carbon
acquisition and allocation (Costes et al., 2006). Thus improvement of
light penetration within tree canopies has been an objective of
management options for physically constraining plant growth (Lauri,
2002). The optimization of tree architecture could also be achieved
through genetics and breeding to complement and eventually reduce human
intervention with training systems. Ideally, the within-species genetic
variability could be used for defining ideotypes in plant breeding.
However, it remains difficult to integrate architectural traits in
breeding programs due to the complex changes in trait values during tree
development (Laurens et al., 2000).

\subsubsection{Model description}\label{model-description}

There exist few methods to quantify and objectively compare the impact
of training systems and cultivars on light interception efficiency.
Moreover, the complexity of fruit tree structure, the large number of
trees required for experiments in quantitative genetics, and the long
growth period makes it difficult to use real trees to explore the link
between the genetic variation of tree architecture and light
interception throughout tree development.

In this context, MAppleT was developed to explore wide ranges of tree
geometries and topologies \emph{in silico}. MAppleT is a
functional-structural plant model that simulates apple tree development
over years, considering both topology and geometry in interaction with
the environment (including gravity in regard to branch bending) (Costes
et al., 2008). The growth and branching processes are simulated with
Markov chains and hidden semi-Markov chains, respectively, estimated
based on previously collected data (Costes and Guédon, 2002; Costes et
al., 2003; Renton et al., 2006). For geometrical development, branch
bending is supported by a biomechanical model (Alméras, 2001;
Taylor-Hell, 2005), taking into account the intra-year dynamics of
primary, secondary and fruit growth. A detailed documentation on the
model variables, parameters, and implementation can be found online at
the MAppleT project repository (Cokelaer, 2017).

Initially parameterized for the Fuji cultivar, MAppleT outputs represent
the progression of tree form and topology over time and were assessed by
comparing descriptors between simulated and digitized trees (Costes et
al., 2008). The model was used in virtual experiments in which apple
trees were simulated and coupled with \(\mu\)SLIM, i.e.~MultiScale Light
Interception Model, which estimates radiation attenuation from
statistical description of foliage at different scales (Da Silva et al.,
2008). \emph{In silico} experiments explored different combinations of
geometrical and topological traits and their effects on light
interception, considered as one of the key parameters to optimize fruit
tree production (D. Da Silva et al., 2014; David Da Silva et al., 2014;
Han et al., 2012).

\subsubsection{Design of experiments}\label{design-of-experiments}

Our work focused on four key geometrical traits: branching angle (BA),
internode length (IL), top shoot diameter (TSD), and leaf area (LA).
These traits were selected based on previous investigations by (D. Da
Silva et al., 2014; Han et al., 2012) because of the assumed influence
on the three-dimensional distribution of leaves and therefore on light
interception. The performance metric in this study is the integrated
projected leaf area (iPLA) for five-year-old trees, a global descriptor
of the light interception efficiency of a tree. The iPLA is the result
of a weighted mean of 46 directional projected leaf areas. The 46 used
directions are the central directions of 46 solid angle sectors of equal
area discretizing the sky hemisphere according to the `Turtle sky'
proposed by (Dulk, 1989). Each direction is associated with a weighting
coefficient related to its elevation, the lower the elevation, the
smaller the coefficient as is the sun radiance.

Since petiole angles are constant in the current version of MAppleT, the
leaf orientation was assumed to be primarily influenced by branching
angle and branch bending. The latter depends, for a given wood
elasticity, on the weights imposed by leaves and internode widths along
an axis. The internode widths are recursively accumulated from the shoot
top to the shoot base. Thus, the top shoot diameter is expected to have
an impact on branch bending and consequently on leaf spatial
distribution. Internode length, which determines the distance between
leaves, may impact leaf density and branching. Finally, the area of each
individual leaf is, along with the number of leaves, a major component
for total interception surface of the tree. However, not all leaf area
captures light and the projected leaf area accounts for overlaps and
mutual shading of leaves in the canopy. \(iPLA\) is thus considered as a
proxy of the intercepted light by the whole tree.

Topological variables were not considered in this study. However,
because we used Markov chains, the topology was not constant over all
geometries. To avoid drawing conclusions specific to a particular
configuration, the performance is taken as the mean over three distinct
topologies, obtained from given values of the four input parameters. In
summary, denoting \(t_1, t_2, t_3\) the topologies, the performance can
be defined as:

\begin{equation}
 P(\x) = \frac{1}{3} \sum_{i=1}^3 iPLA(\x, t_i),
\end{equation}

with \(\x = \{BA, IL, TSD, LA \}\) being the set of values of input
parameters.

The genotype-dependent parameters were obtained by measuring 123 apple
tree hybrids developed from the biparental cross between
\emph{Starkrimson} and \emph{Granny Smith}. These trees were planted and
measured during several years in an experiment reported by Segura et al.
(2008). Briefly, 246 trees (two replicates) were planted in March 2004
at the Melgueil INRA Montpellier experimental station 5 x 1.5 m apart in
an east--west orientation in six-tree microplots randomly scattered
throughout the field. All the trees were grown with minimal training,
that is, they were not pruned and the trunks were staked up to 1 m. They
were regularly irrigated using a microjet system to avoid soil water
deficits. Pests and diseases were controlled by conventional means in
line with professional practices throughout this study. This dataset
provides 6,150 average measurements from the individual trees for the
four geometrical traits. The interval of variation for each trait is
reported in Table 1. The minimum and maximum values for the four studied
parameters correspond to the minimum and maximum values observed within
the progeny. For reference, those extremes can be compared to the
parameters for the Fuji cultivar: 45 for BA, 0.03 for IL, 0.003 for TSD
and LA.

\begin{quote}
\textbf{Table 1: Description and variation range for input parameters of
the MAppleT model.}
\end{quote}

\subsection{Sunflower}\label{sunflower}

Sunflower is adapted to a wide range of environmental conditions (de la
Vega and Hall, 2002) and breeders mainly focus on productivity, disease
resistance and fatty acid composition, while targeting large climatic
zones within geographical Europe (including Ukraine and Russia) or
Argentina (de la Vega and Chapman, 2006). Variability in the duration of
vegetative growth and grain filling is the main lever to adapt the
cultivar to farming environments, primarily by growing earlier maturing
cultivars in colder zones. Overall, this broad-adaptation strategy is
also more economical for the breeding industry since it reduces the
number of cultivars to be handled in seed production and distribution
networks.

Sunflower grown in Southern Europe is mostly cultivated in low rainfall
areas, without irrigation, and on shallow soils (Tuck et al., 2006) that
result in frequent exposure to water deficits (Olesen and Bindi, 2002).
Drought tolerance involves a wide range of component processes and their
spatial and temporal combination (Jones, 2007). Consequently, different
phenotypes can support diverse drought adaptation strategies. For
example, minimization of water loss can be achieved by lowering either
leaf area, transpiration per unit leaf area (stomatal conductance) or
reducing the energy load of the plant (extinction coefficient) (Sadras
et al., 1993). The regulation of stomatal conductance, as a function of
water deficit, was found to present genotypic variability (Casadebaig et
al., 2008) and to be a strong determinant of crop productivity under
drought (Casadebaig et al., 2011).

\subsubsection{Model description}\label{model-description-1}

SUNFLO is a process-based model for sunflower that was developed to
simulate the grain yield and oil concentration as a function of time,
environment (soil and climate), management practice and genetic
diversity (Casadebaig et al., 2011; Debaeke et al., 2010; Lecoeur et
al., 2011). This model is based on a conceptual framework initially
proposed by Monteith (1977) and now shared by a large family of crop
models. In this framework, the daily crop dry biomass is calculated as a
difference equation function of incident photosynthetically active
radiation, radiation interception efficiency (RIE) and radiation use
efficiency (RUE, g MJ\textsuperscript{-1}). The radiation interception
efficiency is a function of leaf area index (LAI) and light extinction
coefficient (k), based on Beer-Lambert's law
(\(RIE = 1 - exp^{-k~LAI}\)). The RUE concept (Monteith, 1994) is used
to represent photosynthesis at the crop scale.

Broad scale developmental or physiological processes of this framework,
the dynamics of LAI, photosynthesis (RUE) and biomass allocation to
grains were split into finer processes (e.g leaf expansion and
senescence, response functions to environmental stresses) to account for
genotypic specificity, thus exhibiting G \(\times\) E interactions.
Globally, the SUNFLO crop model has about 50 equations and 64 parameters
(43 plant-related traits, among which eight are genotype-dependent and
21 environment-related). In cropping conditions, these physiological
processes are affected by numerous abiotic or biotic factors. Therefore,
predictions with the SUNFLO model are restricted to obtainable yield
(Van Ittersum and Rabbinge, 1997): only the main limiting abiotic
factors (temperature, light, water and nitrogen) were considered. A
report that summarizes the equations and parameters used in the model is
available as supplementary information. The source code is available on
INRA software repository
(\href{http://mulcyber.toulouse.inra.fr/anonscm/git/sunrise/sunrise.git}{git},
tag 1.4, commit SHA \texttt{4d5c30d3}) and the VLE-RECORD environment
(Bergez et al., 2013; Quesnel et al., 2009) is used as simulation
platform.

SUNFLO was evaluated on both specific research trials (40 trials, 110
plots) and agricultural extension trials that were representative of its
targeted use (96 trials, 888 plots). Over these two datasets, the model
was able to simulate significant G \(\times\) G interaction and rank
genotypes (Casadebaig et al., 2011; Casadebaig, Mestries, et al., 2016).
The prediction error for grain yield was 15.7\% when estimated over all
data (9\% - 30\% in individual trials). From these two evaluations, it
was determined that SUNFLO is accurate enough to support optimization
methods, i.e.~allows discrimination between two given cultivars.

\subsubsection{Design of experiments}\label{design-of-experiments-1}

Previous studies suggested that high yields could be obtained by several
specific trait combinations (Casadebaig and Debaeke, 2011; Lecoeur et
al., 2011). SUNFLO inputs include genotype, environment (soil and
climate), management actions, and initial conditions. In this study
eight genotype-dependent input parameters were included for optimization
(Table 2) whereas management actions (sowing on 2012-04-16, 7 plants
m\textsuperscript{-2}, no irrigation nor fertilization), soil
characteristics (150 mm soil water capacity), and initial conditions
(full initial soil water capacity, 30 kg ha\textsuperscript{-1} residual
mineral nitrogen) were fixed. These conditions are representative of the
average low-input farming operations for the sunflower crop in France
(Barbet-Massin, 2011).

Performance was defined as the mean grain yield over five model
evaluations. To capture basic climate variability in these models, data
was collected from five diverse sunflower production regions in France
in 2012 (Avignon, Toulouse, Reims, Poitiers, and Dijon) using
neighboring stations (\textless{} 5km) of the French meteorological
network (Meteo-France).

In summary, denoting \(c_1, c_2, c_3, c_4, c_5\) the five climatic
conditions, the performance can be defined as:

\begin{equation}
 P(\x) = \frac{1}{5} \sum_{i=1}^5 \text{Yield}(\x, c_i),
\end{equation}

with \(\x = \{TDF1, TDM3, TLN, LLH, LLS, K, LE, TR\}\) (see Table 2).

A collection of 89 sunflower hybrid cultivars was phenotyped for
phenology (two traits), architecture (four traits), and response to
abiotic stress (two traits) (Casadebaig et al., 2011; Lecoeur et al.,
2011). The corresponding model parameters are presented in Table 2. The
minimal and maximal bounds were determined for 89 hybrids and assumed to
represent the cultivated genetic diversity.

The values of the genotype-dependent parameters were obtained by
measuring those eight phenotypic traits in dedicated field platforms and
controlled conditions. Our aim was to measure potential trait values, so
different environmental conditions were targeted depending on the set on
traits: field non-limiting conditions (deep soil) for phenological and
architectural traits, field limiting conditions (shallow soil) for
allocation traits, and a range of controlled water deficit (greenhouse)
for response traits. For field experiments, 89 hybrids were phenotyped
on ten trials (two locations, five years: 2008-2012), using randomized
complete block designs with three repetitions of 30 m\textsuperscript{2}
plots (6-7 plant m\textsuperscript{-2}), see Debaeke et al. (2010) and
Casadebaig, Mestries, et al. (2016) for additional details. For
controlled conditions, 82 hybrids were phenotyped in 10 liters pots,
during six greenhouse experiments to determine the response of leaf
expansion and transpiration at the plant scale after stopping watering
and leaving the soil progressively drying (dry-down design). We used
randomized complete block designs with two water treatments (control,
stress) and six repetitions (7 pots m\textsuperscript{-2}), see
Casadebaig et al. (2008) for additional details.

\begin{quote}
\textbf{Table 2: Description and variation range for input parameters of
the SUNFLO model.}
\end{quote}

\subsection{Phenotype optimization}\label{phenotype-optimization}

Phenotype optimization tends to solely focus on performance. Here, we
propose to add a feasibility metric, and to solve a bi-objective
problem.

\subsubsection{Performance}\label{performance}

Phenotypes with the best performance can be identified by solving the
following optimization problem (e.g. Martre, Bénédicte Quilot-Turion, et
al., 2015; Semenov and Stratonovitch, 2013).

\begin{equation}
 \x_P^* = \arg \max_{\x \in \Xset} P(\x), \label{eq:defopt1}
\end{equation}

with \(\x\) representing the phenotype traits and \(P(\x)\) the
performance (respectively, \(iPLA\) and grain yield).
\(\Xset \subset \Rset^d\) defines the ensemble of potential phenotypes
and \(\x_P^*\) the optimal phenotype (ideotype). The dimension of the
problem \(d\) is equal to the number of input parameters, four for
MAppleT and eight for SUNFLO (see Tables 1 and 2).

Choosing \(\Xset\) is, in itself, a difficult task. Typically, one may
define realistic lower and upper bounds for each trait; allowing the
traits to take any value between the bounds. However, solving equation
\ref{eq:defopt1} without any further consideration may lead to
unrealistic (and eventually useless) solutions, as some trait
combinations are very unlikely to be obtained. In particular, extreme
values (or ``corners'') of \(\Xset\), when all traits take their upper
or lower bound values, are highly improbable. Hence, accounting for
correlation between traits, and more generally for the capacity to
obtain a given phenotype, introduces a second objective function, called
\emph{feasibility}.

\subsubsection{Feasibility}\label{feasibility}

Datasets of observed phenotypes were available for both models and used
to define a function that indicates the likelihood of a new phenotype.
These measures provide us with a rough indication of the domain of
potential existence of the traits combinations (both in terms of bounds
and co-occurrence), and a simulated phenotype may be considered less
unrealistic as it is far from the cloud of observations. Therefore, we
require a function that measures the proximity of a simulated phenotype
to the observed phenotypes.

A simple yet sensible solution consists in fitting a multivariate
probability density function to the observations. Such a function would
be maximal at the center of the observation cloud, decrease when moving
away from the center, and naturally take into account the correlations
between traits. Hence, this function could be used as the
\emph{feasibility} function.

As shown in the Results section, \emph{Estimation of feasibility on the
parameter space}, the multivariate Gaussian distribution is a reasonable
choice for our data. Then, using the maximum likelihood estimates for
the distribution moments, the density function is:

\begin{equation}
 \phi(\mathbf{x}) = \frac{1}{\sqrt{(2 \pi)^d | \boldsymbol{\hat \Sigma}|}} \exp \left( -\frac{1}{2} (\mathbf{x} - \mathbf{ \boldsymbol{\hat \mu}})^T \hat \Sigma^{-1}(\mathbf{x} - \mathbf{\boldsymbol{\hat \mu}})  \right),\label{eq:deffeas}
\end{equation}

with \(\boldsymbol{\hat \mu} = \frac{1}{n} \sum_{i=1}^{n} \mathbf{u}_i\)
and
\(\boldsymbol{\hat \Sigma} = \frac{1}{n} \sum_{i=1}^{n} (\mathbf{u}_i- \boldsymbol{\hat \mu}) (\mathbf{u}_i - \boldsymbol{\hat \mu})^T\),
\(n\) being the number of observations (respectively, 123 and 89 in our
use cases) and \(\mathbf{u}_i\) the observed phenotypes.

Finally, for numerical convenience the \emph{feasibility} function is
defined as the logarithm of the density,

\begin{equation}
 F(\x) = \log \phi(\mathbf{x}),
\end{equation}

and the feasibility optimization problem as:

\begin{equation}
 \x_F^* = \arg \max_{\x \in \Xset}  F(\x). \label{eq:defopt1bis}
\end{equation}

\subsubsection{A multi-objective formulation for phenotype
optimization}\label{a-multi-objective-formulation-for-phenotype-optimization}

To identify the most efficient phenotypes, while also favoring those
more likely to be obtained, the initial optimization (equation
\ref{eq:defopt1}) was reformulated as a bi-objective problem:

\begin{equation}
\left\{ \begin{array}{ll}
          \max & \text{Performance} \\
      \max & \text{Feasibility}
        \end{array} \right. \label{eq:defopt2}
\end{equation}

These two objectives are likely to be conflicting, as the most feasible
phenotypes are not necessarily the most efficient ones. So, there will
not exist a common maximizer \(\x^*\) for the two objectives. The goal
is then to identify the set of optimal solutions, called a Pareto set
(Collette and Siarry, 2003), which relies on the concept of Pareto
dominance. A point \emph{dominates} another if both its objectives are
better. Hence, an ensemble of solutions, ranging from the most feasible
to the most efficient is sought.

\subsubsection{Optimization algorithms}\label{optimization-algorithms}

The two models differed substantially in terms of computational need, as
the performance function \(P(\x)\) computation took 0.5s for the SUNFLO
model (annual crop) while it required 135 min for MAppleT (5 year-old
tree), using a 2.90GHz quadcore processor with 8Go RAM. Hence, different
algorithmic families were used to solve the two optimization problems.

For MAppleT, the time cost was prohibitive for using standard
multi-objective optimization algorithms. In a previous study,
metamodel-based optimization strategies were found well-adapted for this
problem with the ability to return a good approximation of the Pareto
set using a very reasonable number of calls to the simulator (Picheny,
2014). In short, this strategy is based on the use of a Gaussian process
approximation model (called \emph{metamodel}, Rasmussen and Williams
(2006)), built from a small number of well-chosen simulations (the
experimental design). This metamodel is then used as a guide to
sequentially choose the most interesting simulations to run (as in the
classical EGO algorithm for single objective problems of Jones et al.,
1998). The optimization process relied on the \texttt{R} statistical
computing environment (R Core Team, 2016) and \texttt{R} package
\texttt{DiceKriging} (v1.5.5, Roustant et al., 2012) and
\texttt{GPareto} (v1.0.3, Binois and Picheny, 2016). 50 initial
simulations were based on a latin hypercube design (McKay et al., 1979)
to build the initial metamodel and 50 simulations were iteratively added
to the experimental design (for a computational time around 24 hours).

For SUNFLO, population-based optimization algorithms that require an
important number of simulations were used, since computational time was
not critical and such algorithms are known to be efficient and reliable
(Collette and Siarry, 2003). Two state-of-the-art algorithms were
tested, namely Non-dominated Sorting Genetic Algorithm-II (NSGA-II, Deb
et al., 2002) and Multi-objective Particle Swarm Optimization with
Crowding Distance (MOPSO-CD, Raquel and Naval, 2005) available in the
\texttt{R} packages \texttt{mco} (v1.0-15.1, Mersmann, 2014) and
\texttt{mopsocd} (v0.5.1, Naval, 2013), respectively. The NSGA-II and
MOPSO-CD algorithms returned slightly different results. Better
convergence was observed with NSGA-II, but wider Pareto front coverage
with MOPSO-CD. Thus, both algorithms were run and the results combined
by extracting all non-dominated solutions from both Pareto sets. This
generated a better overall Pareto front. In these experiments, NSGA-II
was run with 200 generations of 100 individuals and MOPSO-CD was run
with 100 generations of 200 individuals, for a total of 202,000
simulations. These computations had a computational time approximating
nine hours.

\section{Results}\label{results}

\subsection{Estimation of feasibility on the parameter
space}\label{estimation-of-feasibility-on-the-parameter-space}

Graphical representations of the feasibility functions for the MAppleT
and SUNFLO models, along with empirical correlations between traits
(Figure 1) are reported here for the first time. For both models, the
normality hypothesis was observed to be reasonable (except for K and LE
traits in SUNFLO, p \textless{} 0.1), since contour lines match
observations on the projected spaces. The observed variability was
important in all traits, which leads to positive expectations for
selection. The orientation of the ellipsoids shows how correlation
between traits is taken into account in the feasibility estimation.

In apple tree, correlations between the four variables were relatively
low. Despite the usual consideration of allometric relationships between
organ dimensions, especially between internode length and leaf area,
these relationships appeared moderately conserved after genome
recombinations.\\
In sunflower, correlations between traits were globally low, with the
exception of four cases, specifically TDF1 and TDM3; K and TR; TLN and
LLS; TLN and K. Among these exceptions, some observed correlations were
expected, such as the positive correlation in development stages
(earliness at flowering, TDF1 and at maturity, TDM3) or the positive
correlation in plant architecture (high leaf number, TLN and potential
leaf area, LLS). The strong negative correlation between light
extinction coefficient (K) and control of stomatal conductance (TR,
i.e.~how strongly plant transpiration is reduced with water deficit) was
not expected, as these two traits were measured by different methods in
different environments (field versus greenhouse). This correlation
suggests that cultivars that are more efficient at intercepting light
for a given leaf area (high extinction coefficient) are also maintaining
their stomatal conductance under water deficit (high response
parameter).

\includegraphics{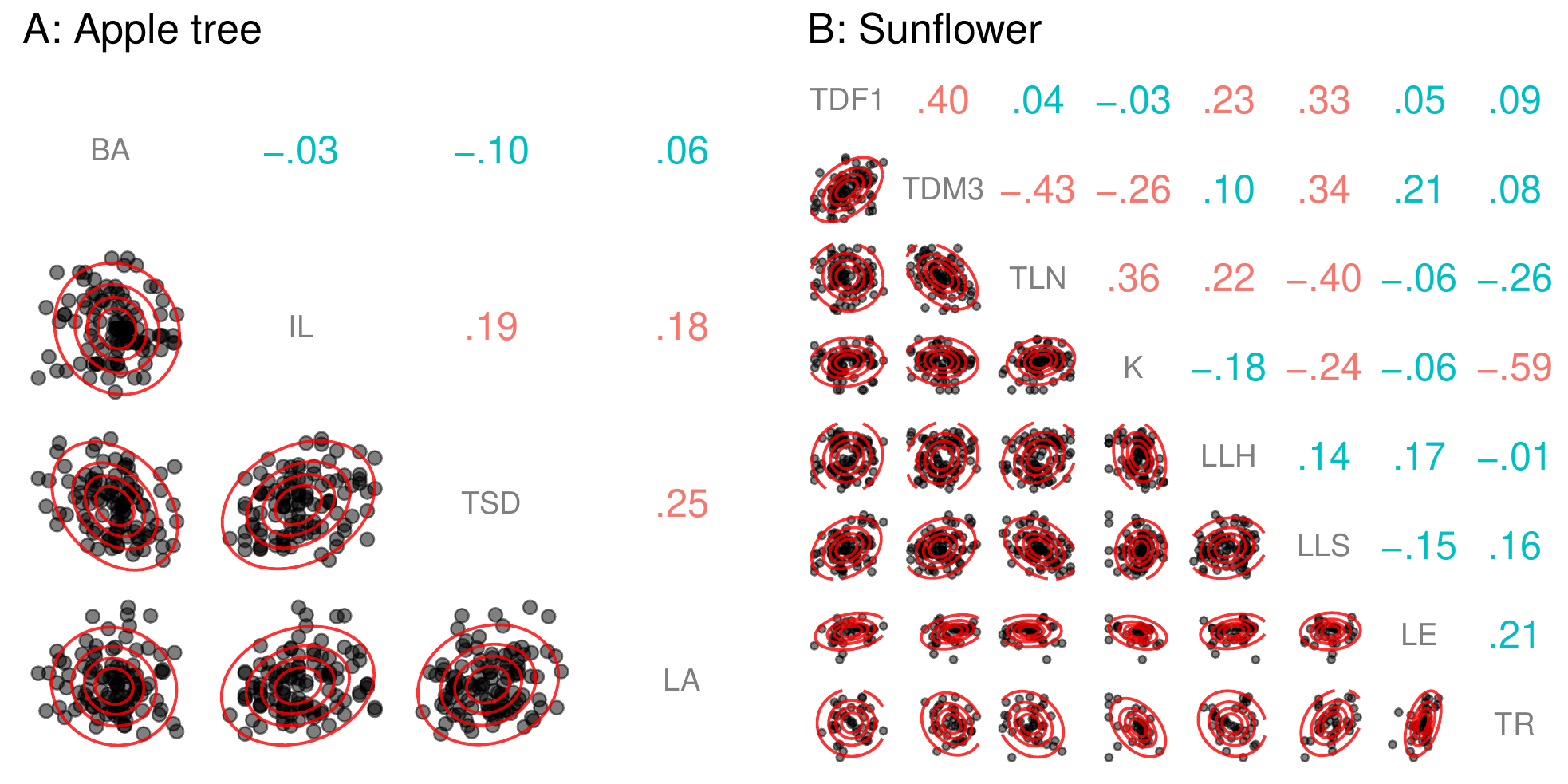}

\begin{quote}
\textbf{Figure 1: Bi-dimensional projections of the observed input
traits and section views of the feasibility function} Dots are
observations made on (A) 123 apple tree hybrids issued from the
biparental cross (Segura et al., 2008) and (B) a collection of 89
commercial sunflower hybrids (n = 42 for LE and TR traits) used as
genotype-dependent parameters. The numbers are the empirical
correlations between traits (red are significant Pearson's \(r\)). The
feasibility function was estimated using a multivariate normal
distribution and was represented with red contour lines (corresponding
to the 25, 50, 75 and 95th percentiles). See Tables 1 and 2 for trait
abbreviation definitions.
\end{quote}

\subsection{Optimization of input
traits}\label{optimization-of-input-traits}

The two Pareto fronts obtained by the optimization algorithms and the
objective values (performance) of (1) the phenotypes corresponding to
the initial experimental design for MAppleT and (2) 50 randomly selected
phenotypes in \([0,1]^d\) for SUNFLO are given in Figure 2. In both
cases, the Pareto fronts were relatively smooth and the first point on
the front (top left) corresponded to the center of the cloud of
phenotypes (i.e.~to the most feasible phenotype). The performance varied
from 60 to 120 for MAppleT, doubling from the most feasible to a very
``atypical'' phenotype. The range of variation in the optimal set is
smaller for SUNFLO, from 2.49 to 2.72 t ha\textsuperscript{-1}, yet the
values of the 50 random phenotypes indicated that the Pareto-optimal
phenotypes were substantially better than average: the mean performance
of the optimal set was around 2.62 t ha\textsuperscript{-1} compared to
2.48 t ha\textsuperscript{-1} for the random set. The most efficient
phenotype (far right on the Pareto front) reached a performance level of
2.72 t ha\textsuperscript{-1}.

\includegraphics{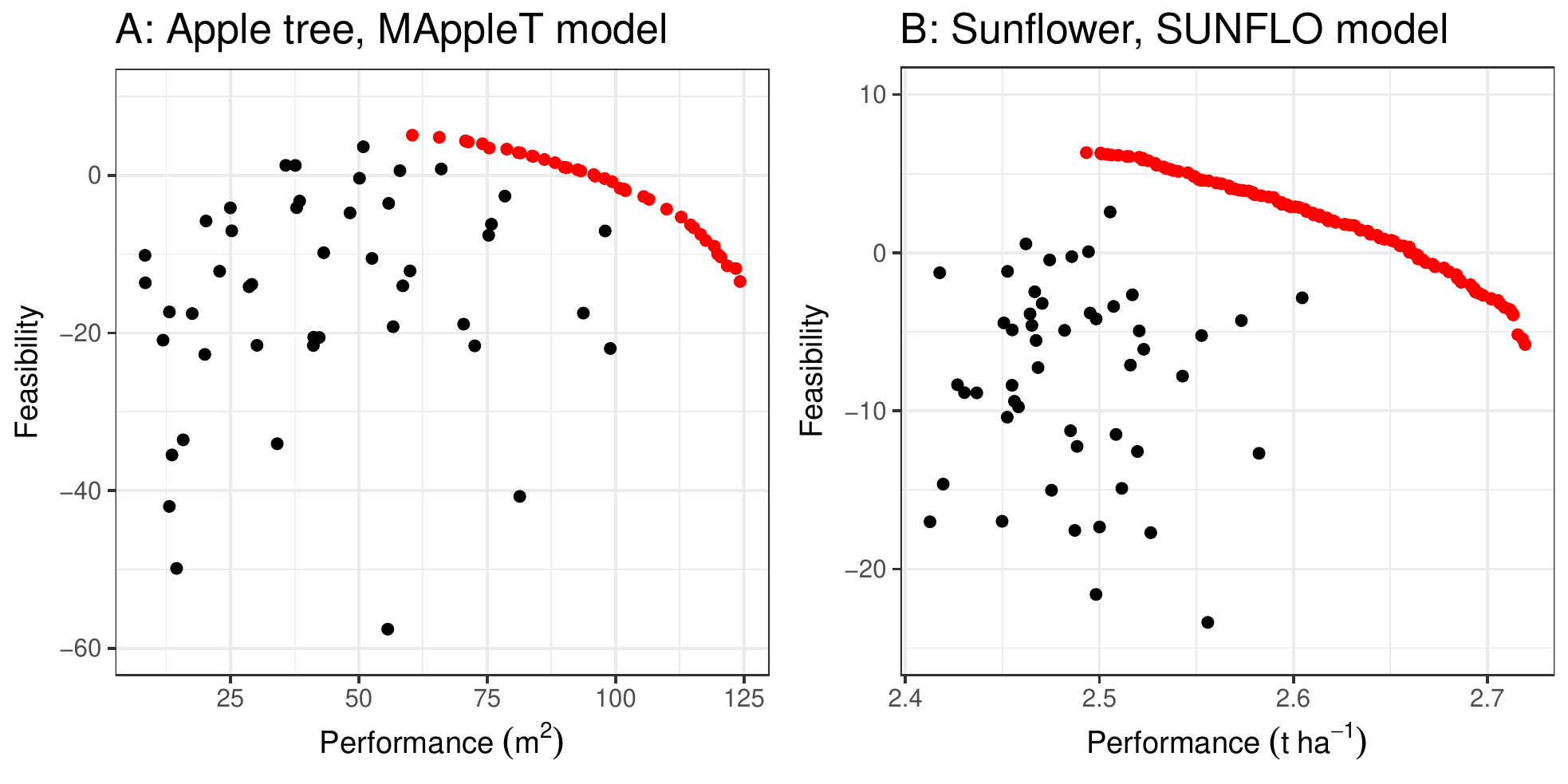}

\begin{quote}
\textbf{Figure 2: Performance and feasibility of sampled and optimized
individuals for the apple and sunflower models.} Optimized individuals
(red dots), which together generate the corresponding Pareto front (red
line) and individuals (black dots) in the initial experimental design
(MAppleT, panel A) or randomly taken in the search space (SUNFLO, panel
B).
\end{quote}

The relationship between the feasibility and performance criteria was
illustrated by displaying the Pareto sets (input values) using
two-dimensional subspace projections (Figure 3). The most central points
are the most feasible, and the points closest to the boundaries are the
most efficient. In both models, the Pareto sets took the form of a
`path', going from the center of the ellipsoids to the bounds of the
hypercube, with the exception of a few traits that remained constant
over the Pareto set (BA in apple tree, TDF1 and LLS in sunflower).

For the apple tree (Figure 3A), the path notably extended outside the
ellipsoids for LA suggesting that efficient phenotypes could exist, but
with very low feasibility. Such phenotypes would rely on a broken
correlation between LA and the three other variables.\\
For sunflower (Figure 3B), the path followed the lowest feasibility
gradient, as indicated by contour lines, where the optimization process
globally followed the correlation between traits. The only notable
exception was for phenology, where the correlation between earliness at
flowering (TDF1) and at maturity (TDM3) was broken in order to reach an
increase in the performance criterion.

\includegraphics{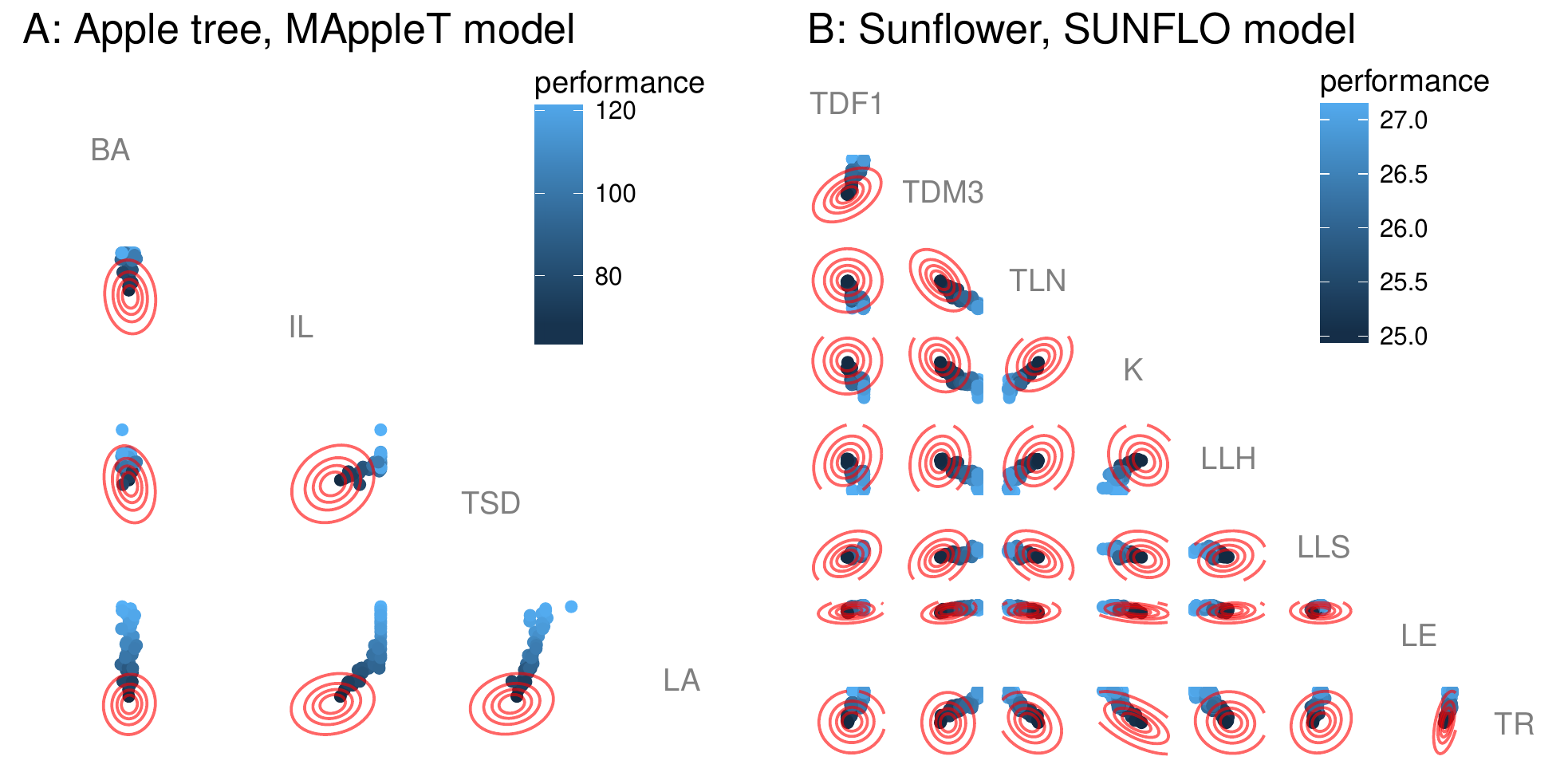}

\begin{quote}
\textbf{Figure 3: Position of optimal phenotypes as a function of
observed trait correlations in sampled cultivars.} The red contour lines
represent the feasibility function: the smaller the ellipse, the larger
the feasibility (cf.~Figure 1). The points show all the Pareto-optimal
phenotypes (optimized individuals). Their intensity of blue color
corresponds to the performance value (in t ha\textsuperscript{-1} for
panel B), with less intense coloration attributed to higher performance.
\end{quote}

To better characterize those paths and the performance:feasibility
trade-off, the value of each trait was plotted for the optimal phenotype
as a function of performance (Figure 4). Since the mean and variance of
each trait were different, we standardized the trait values
(i.e.~substract the mean and divide by the standard deviation of average
individual). Based on this visualization, values exceeding
\(\pm 1.96 \sigma\) were considered as difficult to access based on the
sampled diversity. In addition, we approximated each curve using a
linear model (or piecewise linear model for the IL trait) to improve
readability.

The resultant graph allowed identification of traits whose value was
almost constant in the optimal set of solutions, such as BA for apple
tree (Figure 4A), indicating minimal importance in the
performance:feasibility trade-off. In contrast, lines with larger slopes
corresponded to traits for which the ``price'' to pay for feasibility
was the highest, yet changing their value had the strongest impact on
performance.

For the apple tree (Figure 4A), BA clearly stayed at its mean value,
while IL and LA increased rapidly with performance, reaching atypical
values early. In comparison, the TSD also increased, but at a lower
rate. For sunflower (Figure 4B), the traits could be grouped into sets
of high slopes (TDM3, TR, TLN, K, LLH) or low slopes (TDF1, LE and LLS).
Interestingly, each group contained traits associated with different
plant characteristics, e.g architectural traits controlling leaf profile
(LLH and LLS). These results suggest that the breeding process released
genotypes that perform using diverse physiological strategies, and
opportunities exist for new plant types.

\includegraphics{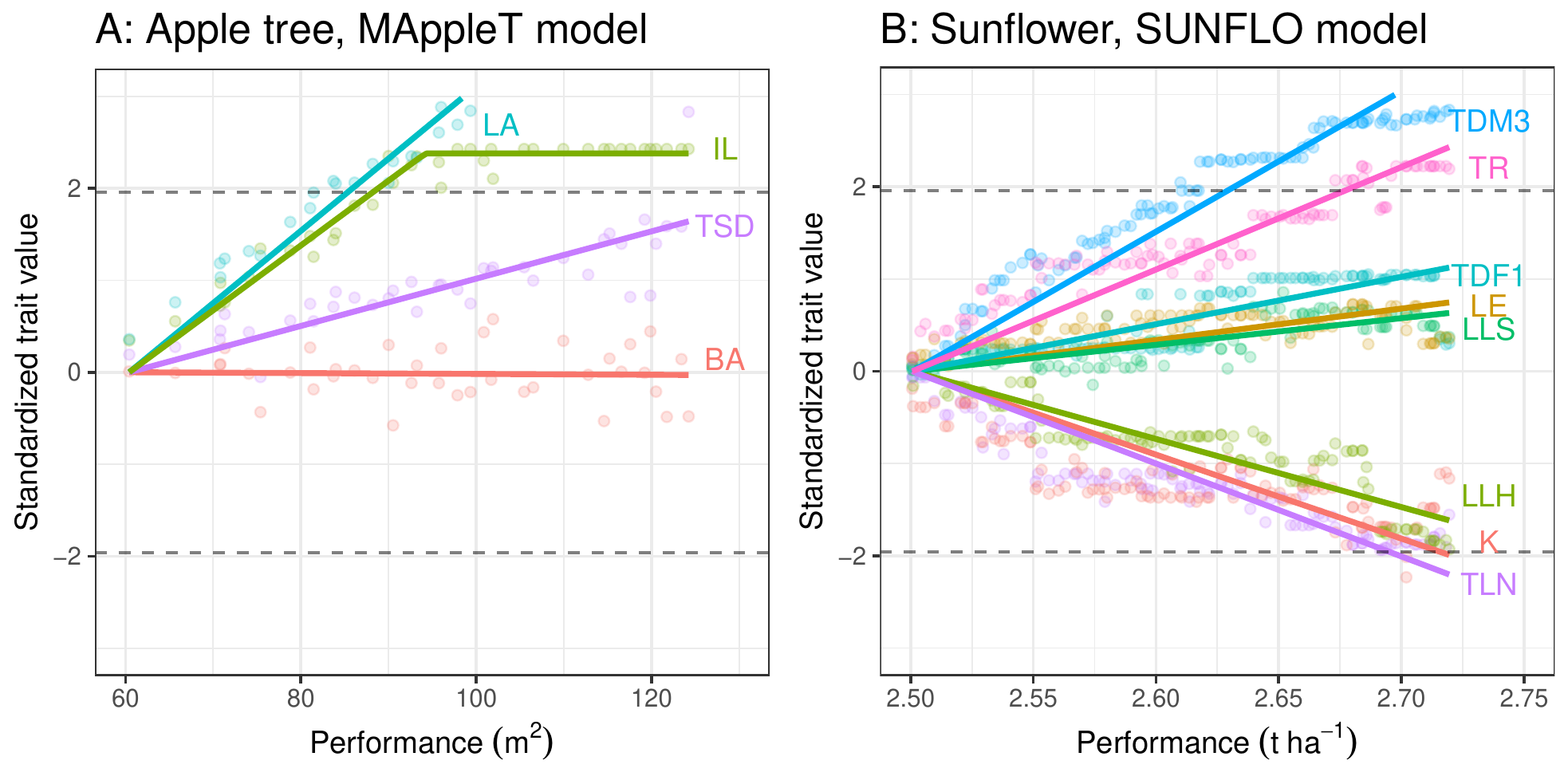}

\begin{quote}
\textbf{Figure 4: The performance:feasibility trade-off in the Pareto
set for apple and sunflower models.} Points represent standardized
traits values as a function of plant performance, for ideotypes in the
Pareto set (n=39 for MappleT, panel A and n=109 for SUNFLO, panel B).
Lines are a linear approximation of the changes in trait values with
increased performance. Dashed lines represent 95\% of available
variability (\(\pm 1.96~\sigma\)), i.e.~trait values exceeding this
threshold can be considered as difficult to access. See Tables 1 and 2
for trait abbreviation definitions.
\end{quote}

Finally, we reported four optimal plant phenotypes evenly distributed
along the Pareto front for apple tree and sunflower (Table 3).

\begin{quote}
\textbf{Table 3: Eight optimal phenotypes identified with the MappleT
and SUNFLO models.}
\end{quote}

\section{Discussion}\label{discussion}

The use of phenotypic measurements of existing cultivar populations and
simulation-based optimization allowed us to virtually recombine plant
traits to identify a set of optimal phenotypes that maximize performance
and feasibility (Figure 2, Tables 3-4). This result is first discussed
from a biological perspective, i.e the morphological and physiological
characteristics in the optimal set of phenotypes for apple tree and
sunflower. We then discuss the influence of the input dataset on the
outcome of our approach and how the feasibility criterion could be
defined in relation to the type of genetic material (e.g recombinant
inbred lines or hybrids). Finally, we discuss how this multi-objective
optimization method is a less data-intensive alternative than emergent
molecular breeding methods such as genomic selection or gene-based
modeling to estimate the breeding value of plant phenotypes.

\subsection{Morphological and physiological characteristics of the
optimized
phenotypes}\label{morphological-and-physiological-characteristics-of-the-optimized-phenotypes}

Considering both apple and sunflower use cases, the
performance:feasibility optimization approach resulted in paths of
improvement that can be used in trait-based breeding (Figure 4A and 4B).
Such paths could be used to follow target values for the most promising
traits. These paths also use trait covariance to indicate when
modifications would yield an unlikely plant phenotype.

\subsubsection{Apple tree use case: longer internodes and higher leaf
areas improve light
interception}\label{apple-tree-use-case-longer-internodes-and-higher-leaf-areas-improve-light-interception}

Branching angle (BA) in apple trees can range from 20 to 120 degrees
(Table 1), but values approximately 80° from the vertical (open
branches) were identified as both the most feasible and most optimal
(Table 3). Morever, due to branch bending, this basal branching angle
will result in higher top branching angles. However, no significant
improvement in this trait can be expected, since there was no increase
in the performance value associated with its variation (Figures 4A and
4B). This result is consistent with observations in more erect trees
with low branch angles where light capture efficiency is reduced, likely
from higher leaf overlap (David Da Silva et al., 2014).

In contrast, significant increases in performance could be expected by
increasing internode length, leaf area, and to a lesser extent top shoot
diameter (Table 3, Figure 4A). Interestingly, the highest value in the
explored range of internode length was reached prior to maximum
performance suggesting that internodes longer than five cm could
increase \(iPLA\) and therefore light interception. However, this
extreme phenotype is less feasible than other phenotypes, as well as
unstable from a biomechanics point of view; the corresponding shoots
would likely be unable to support the weight of the apple fruit (Alméras
and Fournier, 2009).\\
The existence of such tradeoffs among traits has been shown in other
species, especially long internodes increasing light interception but
being detrimental to biomechanics (Pearcy et al., 2005). These tradeoffs
are likely to result from developmental, biomechanical and hydraulic
constraints in plants and to be responsible for the general rules
observed across species between stem diameter and leaf area (Corner,
1949), as well as for limits in plant plasticity (Valladares et al.,
2007).

No limitation was observed for leaf area or top shoot diameter where
increases continued to be associated with higher performance (Figure
4B).These results suggest that values beyond the scope of this work
could be explored for increased light interception, with the caveat that
such phenotypes would be associated with low feasibility (Table 3).
Performance advantages from increased top shoot diameter were less rapid
than with leaf area. This result further suggests that exploring values
beyond those tested here present additional opportunities for
improvement of performance, at least in terms of light interception.\\
As previously mentioned for internode, tradeoffs may exist between leaf
structure and its functions. Leaf morphology and specific weight, which
depend on the intercepted light during organogenesis, are tightly linked
to leaf functions, photosynthesis and transpiration. Moreover, whatever
the trait considered, dimensions are correlated to biomass and organ
construction and maintenance have a cost that may counter-balance the
expected advantages (H. Poorter et al., 2006). Even though the space of
correlation values explored here was less than in ecological studies
(e.g. Niklas, 1994; L. Poorter et al., 2006), both constraints and costs
are likely to limit plant plasticity and the domain of possible
phenotypes (Valladares et al., 2007). Accounting for leaf respiration
and transpiration would certainly limit the performance of plants with
large leaf areas. The fact that no plants were observed with such large
leaf areas could mean they are sub-optimal. Making models more complex,
including more constraints among traits and criteria to define
optimality would help detecting such trade-off but remain challenging.

The optimal phenotypes designed in the present approach exhibit
characteristics quite different from those of the Fuji variety used in
the initial version of MAppleT (Costes et al., 2008; David Da Silva et
al., 2014). Indeed, longer internodes, and higher leaf areas would be
preferable for optimizing light interception.

\subsubsection{Sunflower crop use case: late maturation, lower leaf area
and adaptive traits improve
performance}\label{sunflower-crop-use-case-late-maturation-lower-leaf-area-and-adaptive-traits-improve-performance}

The values for two sunflower leaf traits, specifically LLS and LE,
remained relatively constant across the optimal set of solutions,
indicating that once fixed to their optimal values, these traits had no
impact in the performance:feasibility trade-off (Figure 4B). The
designed sunflower ideotype, for the considered growth environments, had
a less than average potential individual leaf area (low LLS in Table 3)
and leaf expansion was more sensitive to water deficit (high LE in Table
3), compared to the overall population of phenotyped hybrids (Table 2).
For these two traits, the convergence towards a fixed value indicated
that they were central for cultivar adaptation in the sampled
environments: any modifications in these traits would lead to
sub-optimal solutions (either in performance or feasibility).

In contrast, modification of the other six traits had an impact on the
performance:feasibility trade-off of the ideotypes (i.e.~important
slopes in Figure 4B; Table 3). The most efficient ideotype had a late
maturity date (TDM3) and mid-late flowering date (TDF1). Its aerial
architecture was defined by a low leaf number (TLN), a symmetric leaf
profile (LLH), and a low extinction coefficient (K). Sensitive control
of the stomatal conductance (TR), where stomatal conductance starts to
decline at moderate water deficits, was also found to be an important
characteristic. These optimal characteristics were difficult to obtain,
at least for maturity and control of stomatal conductance, as the values
were out of the \([-2,+2]\) feasibility range.

Interestingly, the characteristics of the designed ideotype portrayed a
plant well-adapted to water deficit: moderate leaf area, low light
extinction, and possessing adaptive traits (LE, TR). These
characteristics lead to water-efficient plants, with a lower water loss
due to transpiration, because the leaf area is lower or because
transpiration rate is reduced under water deficit (higher LE value).

\subsection{Improving genotype-to-phenotype prediction in
simulation-based ideotype
design}\label{improving-genotype-to-phenotype-prediction-in-simulation-based-ideotype-design}

\textbf{Genericity and limits of the results depend on model complexity
and targeted cropping environments} In our experiments, the input
phenotypic space was bounded by observed values, thereby avoiding the
question of extrapolating plant performances for unseen-before traits
values. Although our approach indicates ``directions'' for ideotypes,
extrapolation should only be done with caution. For example, Haile et
al. (1998) and Srinivasan et al. (2017) report empirical evidences that
leaf area is already supra-optimal for the soybean crop (see also Conley
et al., 2008, 2009), whereas our approach indicates a positive value of
increasing leaf area-related inputs (Figure 4, strong value for apple
trees, lesser for the sunflower crop).

This discrepancy can first result from the model content and validity
domain. Plant models are a simplified description of the real biological
system, and some physiological processes might not be represented (e.g
respiration, biotic stresses) leading to an erroneous simulated cost for
biomass production and thus plant performance. We argue that making
models complex enough to detect such trade-offs might be much more
difficult than operating optimization in a bounded phenotypic input
space and incorporating a measure of feasibility.

The definition of cropping environments in the model-based approach can
also lead to differences in trait value. In our approach, average
cropping conditions only were used. Considering additional diverse
environmental and management factors (for instance, drought-prone
environments) led to negative value for leaf area increase (Picheny et
al., 2017). However, considering additional diverse environmental and
management factors (for instance, drought-prone environments) would
likely affect the characteristics of the optimal set of ideotypes (e.g
negative value for leaf area increase as observed in Picheny et al.
(2017)) and eventually lead to ideotypes for specific growth conditions.
We suggest that a few key traits are thus responsible for cultivar
global adaptation capacity. Secondary traits are needed to adapt to
uncertain environmental conditions and support alternative resource use
strategies.

\textbf{Is the feasibility function a reasonable representation of the
biologic reality?} The feasibility function is entirely dependent on the
dataset at hand. This implies that the diversity of optimal phenotypes
selected by the optimization process results from the trait variance
present in the considered population. This variance corresponds to
heterosis (biparental progeny), in which case it also depends on chosen
parents, or to genetic variability (collection of hybrids). When more
polymorphisms are available, with multiple alleles at a locus, more
diverse phenotypic traits can be obtained. The genetic distance between
the parents will also impact this capacity; more distantly related
parents increase the chance of different alleles at a locus.

In this study, the variation observed in the apple tree case resulted
from a biparental segregating population with low parental relatedness
(Allard et al., 2016). A larger distribution of the traits, and possibly
different correlations could be expected from different crosses or in
populations with increased genetic diversity, such as multi-parental
populations (Bink et al., 2002; Blanc et al., 2006) or core collections
(Lassois et al., 2016). In the sunflower case, the observed variability
in the traits resulted from the upstream selection of the commercial
hybrids. The variability observed in this material likely covers only a
small portion of the diversity present in core collections. The lack of
kinship information on this material assumes average values for each
trait are most feasible. This is reflected by our definition of the
feasibility function: maximal at the center of the cloud and lowest at
the edges of the domain. Using plant material with a more complex
genetic structure would necessitate accounting for kinship matrices
comparable to pedigree-based (Bink et al., 2002) or genome-wide
association study (GWAS) analyses. This would account for genetic
correlations between individuals and between traits as proposed
herewith.

More specifically, this approach could consider alternative feasibility
functions depending on the type of genetic material screened in the
optimization process. In the proposed function (Equation 5), the
proximity of a virtual phenotype to a particular existing individual
will only indirectly be accounted for. In the sunflower case, where
hybrids were produced by breeding, an alternative feasibility function
could be defined that would reach its maximum at those particular trait
combinations (Figure 5). In this case, virtual phenotypes matching
existing phenotypes would be more present in optimal solutions, even if
they are farther from the population mean. We did not conduct these
tests because it was not central to our proof of concept study. However,
the global approach and optimization algorithm would support alternative
feasibility functions without modifications.

\includegraphics{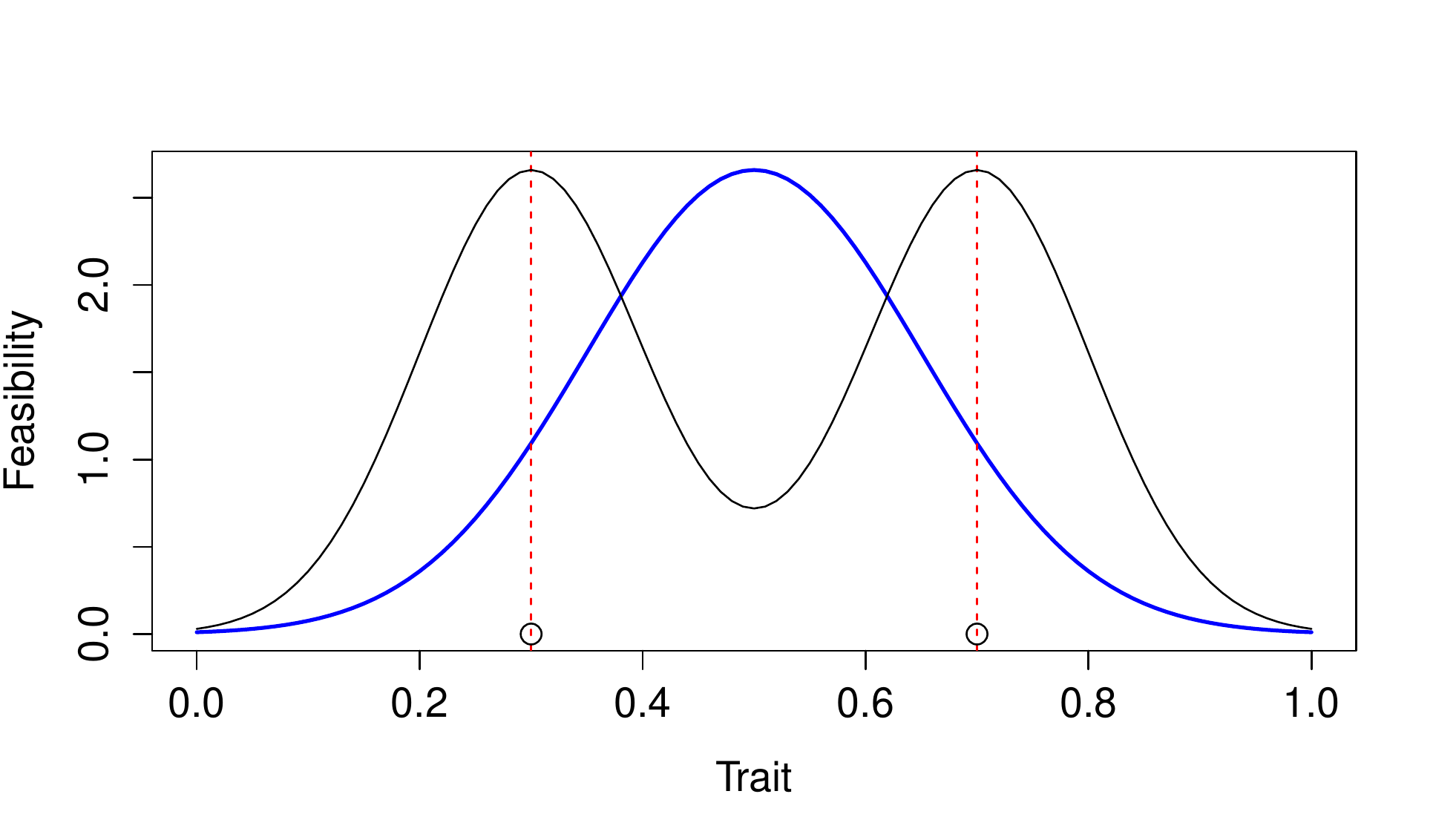}

\begin{quote}
\textbf{Figure 5: Feasibility function used in the study (blue) and
alternative function (black).} The two circles and vertical lines
correspond to existing phenotypes. The assumption behind the current
feasibility function is that it is easier to produce a phenotype with
average traits, while the alternative assumption would be that it is
easier to produce a phenotype close to a single existing one.
\end{quote}

\textbf{How does our approach fit into gene-to-phenotype prediction
strategies ?} The efficiency of the breeding process benefits from
gene-to-phenotype predictions to assign an accurate breeding value to
genotypes. Hybrid modeling approaches have recently emerged between
statistical modeling (\emph{forward}, gene to phenotype) and
process-based modeling (\emph{reverse}, phenotype to gene) (Bustos-Korts
et al., 2016).

On the one hand, gene-to-phenotype predictions are generally undertaken
from the molecular level where genomic selection (Meuwissen et al.,
2001) is used to predict the breeding value from genome-wide
informations and statistical modeling. The difficulty with this method
lies in the prediction of non-additive gene effect and gene \(\times\)
environment interactions. These predictions can stem from environmental
covariables (Heslot et al., 2014) or computed from simulation modeling
(Technow et al., 2015). Additionally, as in our approach, phenotypes
measured in the training population for genomic selection could also be
used to estimate feasibility function based on correlations observed
between traits. This function will likely improve the accuracy of the
genomic selection model for estimating breeding values in the next
generation (e.g. Marulanda et al., 2015).

On the other hand, gene-based modeling (White and Hoogenboom, 2003; Yin
et al., 2004), where gene or QTL action is represented through linear
effects of specific alleles on crop simulation model parameters (Chenu
et al., 2008; Messina et al., 2006), intrinsically accounts for
feasibility at the expense of an intensive phenotyping (i.e.~on large
populations) and modeling process (i.e.~to develop genetic models for
all influent genotype-dependent parameters of the simulation model).

Consequently, gene-to-phenotype predictions can be improved to account
for gene \(\times\) environment interactions by (1) augmenting genomic
selection frameworks with new types of predictors or (2) relying on
numerical models and optimization methods to identify optimal
combinations of traits (or alleles, depending on the input type). Our
approach fits the second option, and can improve the realism of
ideotypes designed with model-based optimization methods. This requires
less phenotyping work as compared to the gene-based modeling options, by
estimating the feasibility criteria using the values of
genotype-dependent parameters of the simulation model.

\section{Conclusion}\label{conclusion}

Plant numerical models are powerful tools to facilitate ideotype design.
These models offer the ability to perform large-scale experiments and to
explore a large variety of trait combinations. However, as stated by
Martre, Bénédicte Quilot-Turion, et al. (2015), it is essential to
integrate genetic constraints on virtual cultivars during trait
optimization to avoid impossible phenotypes that breeders cannot
produce. We argue that an efficient way to increase realism in
model-based ideotype design approaches is by introducing correlations
between genotype-dependent parameters into a feasibility criterion,
which is not formally considered in current approaches. Multi-objective
optimization allowed exploration of potential phenotypes that
successfully target realistic trait combinations and avoid misleading
solutions. This feasibility criterion approach successfully demonstrated
that it could provide paths for desirable and realistic trait
modifications (both in direction and intensity), when coupled with
trait-based breeding methods, for apple trees and sunflower crop in our
proof of concept study.

\section{Acknowledgements}\label{acknowledgements}

The authors are grateful to the students (Ewen Gery, Bertrand Haquin,
Claire Barbet-Massin) and staff from INRA and ENSAT (Colette Quinquiry,
Michel Labarrère, Pierre Maury) and \emph{Terres Inovia} (Pascal Fauvin,
Philippe Christante, André Estragnat, Emmanuelle Mestries) that helped
to constitute the phenotypic database, helped in modeling and simulation
(RECORD team from INRA, Helène Raynal, Eric Casellas, Gauthier Quesnel)
and provided climatic datasets (AgroClim team from INRA). We also thanks
V. Segura, S. Martinez and G. Garcia for apple tree phenotyping, the
OpenAlea team for their support in model development and L. Han for the
first step of sensitivity analyses. We are also grateful to Jennifer
Mach and Shawn Carlson for their insightful inputs while editing the
manuscript. Grants were provided by INRA Plant Biology and Breeding
department and the French Ministry of Research (ANR SUNRISE
ANR-11-BTBR-0005).

\newpage 

\section{Table 1}\label{table-1}

\textbf{Table 1: Description and variation range for input parameters of
the MAppleT model.}

\begin{longtable}[]{@{}lllll@{}}
\toprule
Name & Description & Unit & Min & Max\tabularnewline
\midrule
\endhead
BA & Branching angle & \(degree\) & 20 & 120\tabularnewline
IL & Internode length & \(m\) & 0.008 & 0.051\tabularnewline
TSD & Top shoot diameter & \(m\) & 0.001 & 0.0085\tabularnewline
LA & Leaf area & \(m^2\) & 0.0003 & 0.011\tabularnewline
\bottomrule
\end{longtable}

\newpage

\section{Table 2}\label{table-2}

\textbf{Table 2: Description and variation range for input parameters of
the SUNFLO model.}

\begin{longtable}[]{@{}lllrrr@{}}
\toprule
Name & Description & Unit & Mean & Min & Max\tabularnewline
\midrule
\endhead
TDF1 & Temperature sum from emergence to the beginning of flowering &
\(^{\circ}C.d\) & 836.0 & 765.0 & 907.3\tabularnewline
TDM3 & Temperature sum from emergence to seed physiological maturity &
\(^{\circ}C.d\) & 1673.4 & 1537.6 & 1831.3\tabularnewline
TLN & Potential number of leaves at flowering & \(rank\) & 29.4 & 22.2 &
36.7\tabularnewline
LLH & Potential rank of the plant largest leaf at flowering & \(rank\) &
16.8 & 13.5 & 20.6\tabularnewline
LLS & Potential area of the plant largest leaf at flowering &
\(cm^{-2}\) & 449.9 & 334.3 & 670.0\tabularnewline
K & Light extinction coefficient during vegetative growth & - & 0.9 &
0.8 & 1.0\tabularnewline
LE & Threshold for leaf expansion response to water stress & - & -4.4 &
-15.6 & -2.3\tabularnewline
TR & Threshold for stomatal conductance response to water stress & - &
-9.8 & -14.2 & -5.8\tabularnewline
\bottomrule
\end{longtable}

\newpage

\landscapebegin

\section{Table 3}\label{table-3}

\textbf{Table 3: Eight optimal phenotypes identified with the MappleT
and SUNFLO models.} Each line is an optimal individual sampled from the
Pareto front and ranked by increasing performance. Columns present the
trait combination associated with the plant phenotype.

\begin{longtable}[]{@{}lrrrrrrrrrrrrrrr@{}}
\toprule
Model & Phenotype & BA & IL & TSD & LA & TDF1 & TDM3 & TLN & LLH & LLS &
K & LE & TR & performance & feasibility\tabularnewline
\midrule
\endhead
Apple & 1 & 82 & 0.040 & 0.0060 & 0.0045 & & & & & & & & & 60.00 &
5.08\tabularnewline
& 2 & 81 & 0.048 & 0.0066 & 0.0067 & & & & & & & & & 88.00 &
1.59\tabularnewline
& 3 & 84 & 0.051 & 0.0070 & 0.0089 & & & & & & & & & 110.00 &
-4.30\tabularnewline
& 4 & 77 & 0.051 & 0.0073 & 0.0107 & & & & & & & & & 122.00 &
-11.50\tabularnewline
Sunflower & 1 & & & & & 843.3 & 1693.3 & 28.4 & 17.3 & 478.1 & 0.878 &
-3.76 & -8.82 & 2.52 & 6.03\tabularnewline
& 2 & & & & & 861.8 & 1772.6 & 25.5 & 15.6 & 428.3 & 0.882 & -3.50 &
-7.56 & 2.60 & 3.03\tabularnewline
& 3 & & & & & 866.0 & 1810.7 & 24.2 & 15.7 & 394.7 & 0.894 & -3.63 &
-6.81 & 2.66 & -0.08\tabularnewline
& 4 & & & & & 867.5 & 1828.4 & 23.8 & 15.3 & 341.0 & 0.888 & -3.76 &
-5.88 & 2.71 & -3.87\tabularnewline
\bottomrule
\end{longtable}

\landscapeend

\newpage

\section{Supporting informations}\label{supporting-informations}

\begin{quote}
\textbf{File S1: Conceptual basis, formalisations and parameterization
of the SUNFLO crop model} This document summarises the equations and
parameters used in the SUNFLO crop model.
\end{quote}

\section*{References}\label{references}
\addcontentsline{toc}{section}{References}

\hypertarget{refs}{}
\hypertarget{ref-Allard2016}{}
Allard, A., Bink, M.C., Martinez, S., Kelner, J.-J., Legave, J.-M., Di
Guardo, M., Di Pierro, E.A., Laurens, F., Van de Weg, E.W., Costes, E.,
2016. Detecting qtls and putative candidate genes involved in budbreak
and flowering time in an apple multiparental population. Journal of
experimental botany 67, 2875--2888.
doi:\href{https://doi.org/10.1093/jxb/erw130}{10.1093/jxb/erw130}

\hypertarget{ref-allen2005using}{}
Allen, M., Prusinkiewicz, P., DeJong, T., 2005. Using L-systems for
modeling source--sink interactions, architecture and physiology of
growing trees: The L-PEACH model. New Phytologist 166, 869--880.
doi:\href{https://doi.org/10.1111/j.1469-8137.2005.01348.x}{10.1111/j.1469-8137.2005.01348.x}

\hypertarget{ref-almeras2001acquisition}{}
Alméras, T., 2001. Acquisition de la forme chez des axes ligneux d'un an
de trois variétés d'abricotier: Confrontation de données expérimentales
à un modèle biomécanique (PhD thesis). Agro-Montpellier, Montpellier.

\hypertarget{ref-Almeras2009}{}
Alméras, T., Fournier, M., 2009. Biomechanical design and long-term
stability of trees: Morphological and wood traits involved in the
balance between weight increase and the gravitropic reaction. Journal of
Theoretical Biology 256, 370--381.
doi:\href{https://doi.org/10.1016/j.jtbi.2008.10.011}{10.1016/j.jtbi.2008.10.011}

\hypertarget{ref-Barbet-Massin2011}{}
Barbet-Massin, C., 2011. Quelle représentativité des réseaux
d'homologation variétale et de développement ? Cas du tournesol
(Master's thesis). Ecole d'Ingénieurs Purpan.

\hypertarget{ref-Bergez2013}{}
Bergez, J., Chabrier, P., Gary, C., Jeuffroy, M., Makowski, D., Quesnel,
G., Ramat, E., Raynal, H., Rousse, N., Wallach, D., Debaeke, P., Durand,
P., Duru, M., Dury, J., Faverdin, P., Gascuel-Odoux, C., Garcia, F.,
2013. An open platform to build, evaluate and simulate integrated models
of farming and agro-ecosystems. Environmental Modelling \& Software 39,
39--49.
doi:\href{https://doi.org/10.1016/j.envsoft.2012.03.011}{10.1016/j.envsoft.2012.03.011}

\hypertarget{ref-Bink2002}{}
Bink, M., Uimari, P., Sillanpää, M., Janss, L., Jansen, R., 2002.
Multiple QTL mapping in related plant populations via a
pedigree-analysis approach. Theoretical and Applied Genetics 104,
751--762.
doi:\href{https://doi.org/10.1007/s00122-001-0796-x}{10.1007/s00122-001-0796-x}

\hypertarget{ref-Binois2016}{}
Binois, M., Picheny, V., 2016. GPareto: Gaussian processes for pareto
front estimation and optimization.

\hypertarget{ref-Blanc2006}{}
Blanc, G., Charcosset, A., Mangin, B., Gallais, A., Moreau, L., 2006.
Connected populations for detecting quantitative trait loci and testing
for epistasis: An application in maize. Theoretical and Applied Genetics
113, 206--224.
doi:\href{https://doi.org/10.1007/s00122-006-0287-1}{10.1007/s00122-006-0287-1}

\hypertarget{ref-Brisson2003}{}
Brisson, N., Gary, C., Justes, E., Roche, R., Mary, B., Ripoche, D.,
Zimmer, D., Sierra, J., Bertuzzi, P., Burger, P., Bussière, F.,
Cabidoche, Y.M., Cellier, P., Debaeke, P., Gaudillère, J.P., Hénault,
C., Maraux, F., Seguin, B., Sinoquet, H., 2003. An overview of the crop
model STICS. European Journal of Agronomy 18, 309--332.
doi:\href{https://doi.org/10.1016/s1161-0301(02)00110-7}{10.1016/s1161-0301(02)00110-7}

\hypertarget{ref-Bustos-Korts2016}{}
Bustos-Korts, D., Malosetti, M., Chapman, S., Eeuwijk, F. van, 2016.
Modelling of genotype by environment interaction and prediction of
complex traits across multiple environments as a synthesis of crop
growth modelling, genetics and statistics, in: Crop Systems Biology.
Springer, pp. 55--82.
doi:\href{https://doi.org/10.1007/978-3-319-20562-5_3}{10.1007/978-3-319-20562-5\_3}

\hypertarget{ref-Casadebaig2011a}{}
Casadebaig, P., Debaeke, P., 2011. Using a crop model to assess
genotype-environment interactions in multi-environment trials, in:
Halford, N., Semenov, M. (Eds.), Aspects of Applied Biology, System
Approaches to Crop Improvement. pp. 19--25.

\hypertarget{ref-Casadebaig2008}{}
Casadebaig, P., Debaeke, P., Lecoeur, J., 2008. Thresholds for leaf
expansion and transpiration response to soil water deficit in a range of
sunflower genotypes. European Journal of Agronomy 28, 646--654.
doi:\href{https://doi.org/10.1016/j.eja.2008.02.001}{10.1016/j.eja.2008.02.001}

\hypertarget{ref-Casadebaig2011}{}
Casadebaig, P., Guilioni, L., Lecoeur, J., Christophe, A., Champolivier,
L., Debaeke, P., 2011. SUNFLO, a model to simulate genotype-specific
performance of the sunflower crop in contrasting environments.
Agricultural and Forest Meteorology 151, 163--178.
doi:\href{https://doi.org/10.1016/j.agrformet.2010.09.012}{10.1016/j.agrformet.2010.09.012}

\hypertarget{ref-Casadebaig2016a}{}
Casadebaig, P., Mestries, E., Debaeke, P., 2016. A model-based approach
to assist variety assessment in sunflower crop. European Journal of
Agronomy 81, 92--105.
doi:\href{https://doi.org/10.1016/j.eja.2016.09.001}{10.1016/j.eja.2016.09.001}

\hypertarget{ref-Casadebaig2016}{}
Casadebaig, P., Zheng, B., Chapman, S., Huth, N., Faivre, R., Chenu, K.,
2016. Assessment of the potential impacts of wheat plant traits across
environments by combining crop modeling and global sensitivity analysis.
PLoS ONE 11, e0146385.
doi:\href{https://doi.org/10.1371/journal.pone.0146385}{10.1371/journal.pone.0146385}

\hypertarget{ref-Chapman2003}{}
Chapman, S., Cooper, M., Podlich, D., Hammer, G., 2003. Evaluating Plant
Breeding Strategies by Simulating Gene Action and Dryland Environment
Effects. Agronomy Journal 95, 99--113.
doi:\href{https://doi.org/10.2134/agronj2003.0099}{10.2134/agronj2003.0099}

\hypertarget{ref-Chenu2008}{}
Chenu, K., Chapman, S.C., Hammer, G.L., Mclean, G., Salah, H.B.H.,
Tardieu, F., 2008. Short-term responses of leaf growth rate to water
deficit scale up to whole-plant and crop levels: An integrated modelling
approach in maize. Plant, Cell \& Environment 31, 378--391.
doi:\href{https://doi.org/10.1111/j.1365-3040.2007.01772.x}{10.1111/j.1365-3040.2007.01772.x}

\hypertarget{ref-Cokelaer2017}{}
Cokelaer, T., 2017. MAppleT geference guide {[}WWW Document{]}. URL
\url{http://openalea.gforge.inria.fr/doc/vplants/MAppleT/doc/_build/html/user/autosum.html\#id1}

\hypertarget{ref-collette2003multiobjective}{}
Collette, Y., Siarry, P., 2003. Multiobjective optimization: Principles
and case studies. Springer.

\hypertarget{ref-Conley2008}{}
Conley, S.P., Abendroth, L., Elmore, R., Christmas, E.P., Zarnstorff,
M., 2008. Soybean seed yield and composition response to stand reduction
at vegetative and reproductive stages. Agronomy journal 100, 1666--1669.
doi:\href{https://doi.org/10.2134/agronj2008.0082}{10.2134/agronj2008.0082}

\hypertarget{ref-Conley2009}{}
Conley, S.P., Pedersen, P., Christmas, E.P., 2009. Main-stem node
removal effect on soybean seed yield and composition. Agronomy journal
101, 120--123.
doi:\href{https://doi.org/10.2134/agronj2008.0123}{10.2134/agronj2008.0123}

\hypertarget{ref-Corner1949}{}
Corner, E.J.H., 1949. The durian theory or the origin of the modern
tree. Annals of Botany 13, 367--414.

\hypertarget{ref-costes2002modelling}{}
Costes, E., Guédon, Y., 2002. Modelling branching patterns on 1-year-old
trunks of six apple cultivars. Annals of Botany 89, 513--524.
doi:\href{https://doi.org/10.1093/aob/mcf078}{10.1093/aob/mcf078}

\hypertarget{ref-costes2006analyzing}{}
Costes, E., Lauri, P., Regnard, J., 2006. Analyzing fruit tree
architecture: Implications for tree management and fruit production.
Horticultural reviews 32, 1--61.
doi:\href{https://doi.org/10.1002/9780470767986.ch1}{10.1002/9780470767986.ch1}

\hypertarget{ref-costes2003exploring}{}
Costes, E., Sinoquet, H., Kelner, J.-J., Godin, C., 2003. Exploring
within-tree architectural development of two apple tree cultivars over 6
years. Annals of Botany 91, 91--104.
doi:\href{https://doi.org/10.1093/aob/mcg010}{10.1093/aob/mcg010}

\hypertarget{ref-costes2008mapplet}{}
Costes, E., Smith, C., Renton, M., Guédon, Y., Prusinkiewicz, P., Godin,
C., 2008. MAppleT: Simulation of apple tree development using mixed
stochastic and biomechanical models. Functional Plant Biology 35,
936--950. doi:\href{https://doi.org/10.1071/fp08081}{10.1071/fp08081}

\hypertarget{ref-dasilva2008multiscale}{}
Da Silva, D., Boudon, F., Godin, C., Sinoquet, H., 2008. Multiscale
framework for modeling and analyzing light interception by trees.
Multiscale Modeling \& Simulation 7, 910--933.
doi:\href{https://doi.org/10.1137/08071394x}{10.1137/08071394x}

\hypertarget{ref-dasilva2014light}{}
Da Silva, D., Han, L., Costes, E., 2014. Light interception efficiency
of apple trees: A multiscale computation study based on MAppleT.
Ecological Modelling 290, 45--53.
doi:\href{https://doi.org/10.1016/j.ecolmodel.2013.12.001}{10.1016/j.ecolmodel.2013.12.001}

\hypertarget{ref-dasilva2014investigating}{}
Da Silva, D., Han, L., Faivre, R., Costes, E., 2014. Influence of the
variation of geometrical and topological traits on light interception
efficiency of apple trees: Sensitivity analysis and metamodelling for
ideotype definition. Annals of Botany.
doi:\href{https://doi.org/10.1093/aob/mcu034}{10.1093/aob/mcu034}

\hypertarget{ref-Vega2006}{}
de la Vega, A.J., Chapman, S.C., 2006. Defining Sunflower Selection
Strategies for a Highly Heterogeneous Target Population of Environments.
Crop Science 46, 136--144.
doi:\href{https://doi.org/10.2135/cropsci2005.0170}{10.2135/cropsci2005.0170}

\hypertarget{ref-Vega2002a}{}
de la Vega, A.J., Hall, A.J., 2002. Effects of Planting Date, Genotype,
and Their Interactions on Sunflower Yield: I. Determinants of
Oil-Corrected Grain Yield. Crop Science 42, 1191--1201.

\hypertarget{ref-deb.02}{}
Deb, K., Pratap, A., Agarwal, S., Meyarivan, T., 2002. A fast and
elitist multiobjective genetic algorithm: NSGA-II. Transactions on
Evolutionary Computation 6, 182--197.
doi:\href{https://doi.org/10.1109/4235.996017}{10.1109/4235.996017}

\hypertarget{ref-Debaeke2010}{}
Debaeke, P., Casadebaig, P., Haquin, B., Mestries, E., Palleau, J.-P.,
Salvi, F., 2010. Simulation de la réponse variétale du tournesol à
l'environnement à l'aide du modèle SUNFLO. Oléagineux, Corps Gras,
Lipides 17, 143--51.
doi:\href{https://doi.org/10.1684/ocl.2010.0308}{10.1684/ocl.2010.0308}

\hypertarget{ref-dejong2011using}{}
DeJong, T.M., Da Silva, D., Vos, J., Escobar-Gutiérrez, A.J., 2011.
Using functional--structural plant models to study, understand and
integrate plant development and ecophysiology. Annals of Botany 108,
987--989.

\hypertarget{ref-Ding2016}{}
Ding, W., Xu, L., Wei, Y., Wu, F., Zhu, D., Zhang, Y., Max, N., 2016.
Genetic algorithm based approach to optimize phenotypical traits of
virtual rice. Journal of theoretical biology 403, 59--67.
doi:\href{https://doi.org/10.1016/j.jtbi.2016.05.006}{10.1016/j.jtbi.2016.05.006}

\hypertarget{ref-donald1968breeding}{}
Donald, C.M. t, 1968. The breeding of crop ideotypes. Euphytica 17,
385--403.

\hypertarget{ref-DenDulk1989}{}
Dulk, J.A. den, 1989. The interpretation of remote sensing, a
feasibility study. (PhD thesis). Wageningen Agricultural University.

\hypertarget{ref-edgerton2009increasing}{}
Edgerton, M.D., 2009. Increasing crop productivity to meet global needs
for feed, food, and fuel. Plant physiology 149, 7--13.
doi:\href{https://doi.org/10.1104/pp.108.130195}{10.1104/pp.108.130195}

\hypertarget{ref-Fournier1999}{}
Fournier, C., Andrieu, B., 1999. ADEL-maize: An l-system based model for
the integration of growth processes from the organ to the canopy.
application to regulation of morphogenesis by light availability.
Agronomie 19, 313--327.
doi:\href{https://doi.org/10.1051/agro:19990311}{10.1051/agro:19990311}

\hypertarget{ref-grechi.12}{}
Grechi, I., Ould-Sidi, M.-M., Hilgert, N., Senoussi, R., Sauphanor, B.,
Lescourret, F., 2012. Designing integrated management scenarios using
simulation-based and multi-objective optimization: Application to the
peach tree--Myzus persicae aphid system. Ecological Modelling 246,
47--59.
doi:\href{https://doi.org/10.1016/j.ecolmodel.2012.07.023}{10.1016/j.ecolmodel.2012.07.023}

\hypertarget{ref-Haile1998}{}
Haile, F.J., Higley, L.G., Specht, J.E., Spomer, S.M., 1998. Soybean
leaf morphology and defoliation tolerance. Agronomy Journal 90,
353--362.
doi:\href{https://doi.org/10.2134/agronj1998.00021962009000030007x}{10.2134/agronj1998.00021962009000030007x}

\hypertarget{ref-Hammer2006}{}
Hammer, G., Cooper, M., Tardieu, F., Welch, S., Walsh, B., Eeuwijk, F.
van, Chapman, S., Podlich, D., 2006. Models for navigating biological
complexity in breeding improved crop plants. Trends in Plant Science 11,
587--593.
doi:\href{https://doi.org/10.1016/j.tplants.2006.10.006}{10.1016/j.tplants.2006.10.006}

\hypertarget{ref-han2012investigating}{}
Han, L., Da Silva, D., Boudon, F., Cokelaer, T., Pradal, C., Faivre, R.,
Costes, E., 2012. Investigating the influence of geometrical traits on
light interception efficiency of apple trees: A modelling study with
MAppleT, in: M. Kang, Y. Dumont, and Y. Guo (Eds.), Plant Growth
Modeling, Simulation, Visualization and Their Applications. IEEE Press,
pp. 152--159.
doi:\href{https://doi.org/10.1109/pma.2012.6524827}{10.1109/pma.2012.6524827}

\hypertarget{ref-Heslot2014}{}
Heslot, N., Akdemir, D., Sorrells, M.E., Jannink, J.-L., 2014.
Integrating environmental covariates and crop modeling into the genomic
selection framework to predict genotype by environment interactions.
Theoretical and Applied Genetics 127, 463--480.
doi:\href{https://doi.org/10.1007/s00122-013-2231-5}{10.1007/s00122-013-2231-5}

\hypertarget{ref-Jeuffroy2014}{}
Jeuffroy, M.-H., Casadebaig, P., Debaeke, P., Loyce, C., Meynard, J.-M.,
2014. Agronomic model uses to predict cultivar performance in various
environments and cropping systems. a review. Agronomy for Sustainable
Development 34, 121--137.
doi:\href{https://doi.org/10.1007/s13593-013-0170-9}{10.1007/s13593-013-0170-9}

\hypertarget{ref-jones1998efficient}{}
Jones, D.R., Schonlau, M., Welch, W.J., 1998. Efficient global
optimization of expensive black-box functions. Journal of Global
optimization 13, 455--492.

\hypertarget{ref-Jones2007}{}
Jones, H.G., 2007. Monitoring plant and soil water status: Established
and novel methods revisited and their relevance to studies of drought
tolerance. Journal of Experimental Botany 58, 119--130.
doi:\href{https://doi.org/10.1093/jxb/erl118}{10.1093/jxb/erl118}

\hypertarget{ref-Keating2003}{}
Keating, B.A., Carberry, P.S., Hammer, G.L., Probert, M.E., Robertson,
M.J., Holzworth, D., Huth, N.I., Hargreaves, J.N.G., Meinke, H.,
Hochman, Z., 2003. An overview of APSIM, a model designed for farming
systems simulation. European Journal of Agronomy 18, 267--288.
doi:\href{https://doi.org/10.1016/s1161-0301(02)00108-9}{10.1016/s1161-0301(02)00108-9}

\hypertarget{ref-Lassois2016}{}
Lassois, L., Denancé, C., Ravon, E., Guyader, A., Guisnel, R.,
Hibrand-Saint-Oyant, L., Poncet, C., Lasserre-Zuber, P., Feugey, L.,
Durel, C.-E., 2016. Genetic diversity, population structure, parentage
analysis, and construction of core collections in the french apple
germplasm based on ssr markers. Plant Molecular Biology Reporter 1--18.
doi:\href{https://doi.org/10.1007/s11105-015-0966-7}{10.1007/s11105-015-0966-7}

\hypertarget{ref-laurens2000integration}{}
Laurens, F., Audergon, J., Claverie, J., Duval, H., Germain, E.,
Kervella, J., Lezec, M. le, Lauri, P., Lespinasse, J., others, 2000.
Integration of architectural types in french programmes of ligneous
fruit species genetic improvement. Fruits (Paris) 55, 141--152.

\hypertarget{ref-lauri2002tree}{}
Lauri, P.-É., 2002. From tree architecture to tree training--an overview
of recent concepts developed in apple in france. Journal of the Korean
Society for Horti-cultural Sciences 43 43, 782--788.

\hypertarget{ref-Lecoeur2011}{}
Lecoeur, J., Poiré-Lassus, R., Christophe, A., Pallas, B., Casadebaig,
P., Debaeke, P., Vear, F., Guilioni, L., 2011. Quantifying physiological
determinants of genetic variation for yield potential in sunflower.
SUNFLO: a model-based analysis. Functional Plant Biology 38, 246--259.
doi:\href{https://doi.org/10.1071/fp09189}{10.1071/fp09189}

\hypertarget{ref-lespinasse1992breeding}{}
Lespinasse, Y., Rousselle-Bourgeois, F., Rousselle, P., 1992. Breeding
apple tree: Aims and methods., in: Proceedings of the Joint Conference
of the EAPR Breeding \& Varietal Assessment Section and the EUCARPIA
Potato Section, Landerneau, France, 12-17 January 1992. INRA, pp.
103--110.

\hypertarget{ref-letort.08}{}
Letort, V., Mahe, P., Cournède, P.-H., Reffye, P. de, B., C., 2008.
Quantitative genetics and functional--Structural plant growth models:
Simulation of quantitative trait loci detection for model parameters and
application to potential yield optimization. Annals of Botany 101,
1243--1254.
doi:\href{https://doi.org/10.1093/aob/mcm197}{10.1093/aob/mcm197}

\hypertarget{ref-lopez2008integrating}{}
Lopez, G., Favreau, R.R., Smith, C., Costes, E., Prusinkiewicz, P.,
DeJong, T.M., 2008. Integrating simulation of architectural development
and source--sink behaviour of peach trees by incorporating markov chains
and physiological organ function submodels into L-PEACH. Functional
Plant Biology 35, 761--771.
doi:\href{https://doi.org/10.1071/fp08039}{10.1071/fp08039}

\hypertarget{ref-Martre2015b}{}
Martre, P., He, J., Le Gouis, J., Semenov, M.A., 2015. In silico system
analysis of physiological traits determining grain yield and protein
concentration for wheat as influenced by climate and crop management.
Journal of Experimental Botany.
doi:\href{https://doi.org/10.1093/jxb/erv049}{10.1093/jxb/erv049}

\hypertarget{ref-martre.15}{}
Martre, P., Quilot-Turion, B., Luquet, D., Ould-Sidi Memmah, M., Chenu,
K., Debaeke, P., 2015. Model assisted phenotyping and ideotype design,
in: Sadras, V., Calderini, D. (Eds.), Crop Physiology: Applications for
Genetic Improvement and Agronomy. Academic Press, pp. 349--373.
doi:\href{https://doi.org/10.1016/b978-0-12-417104-6.00014-5}{10.1016/b978-0-12-417104-6.00014-5}

\hypertarget{ref-Marulanda2015}{}
Marulanda, J.J., Melchinger, A.E., Würschum, T., 2015. Genomic selection
in biparental populations: Assessment of parameters for optimum
estimation set design. Plant Breeding 134, 623--630.
doi:\href{https://doi.org/10.1111/pbr.12317}{10.1111/pbr.12317}

\hypertarget{ref-mckay1979comparison}{}
McKay, M.D., Beckman, R.J., Conover, W.J., 1979. Comparison of three
methods for selecting values of input variables in the analysis of
output from a computer code. Technometrics 21, 239--245.

\hypertarget{ref-Mersmann2014}{}
Mersmann, O., 2014. Mco: Multiple criteria optimization algorithms and
related functions.

\hypertarget{ref-Messina2006a}{}
Messina, C., Boote, K., Vallejos, C., others, 2006. A gene-based model
to simulate soybean development and yield responses to environment. Crop
Science 46, 456.
doi:\href{https://doi.org/10.2135/cropsci2005.04-0372}{10.2135/cropsci2005.04-0372}

\hypertarget{ref-Meuwissen2001}{}
Meuwissen, T., Hayes, B., Goddard, M., 2001. Prediction of total genetic
value using genome-wide dense marker maps. Genetics 157, 1819.

\hypertarget{ref-Monteith1994}{}
Monteith, J.L., 1994. Validity of the correlation between intercepted
radiation and biomass. Agricultural and Forest Meteorology 68, 213--220.

\hypertarget{ref-Monteith1977}{}
Monteith, J.L., 1977. Climate and the Efficiency of Crop Production in
Britain. Philosophical Transactions of the Royal Society of London.
Series B, Biological Sciences 281, 277--294.

\hypertarget{ref-Naval2013}{}
Naval, P., 2013. Mopsocd: MOPSOCD: Multi-objective particle swarm
optimization with crowding distance.

\hypertarget{ref-nikinmaa2003shoot}{}
Nikinmaa, E., Messier, C., Sievänen, R., Perttunen, J., Lehtonen, M.,
2003. Shoot growth and crown development: Effect of crown position in
three-dimensional simulations. Tree Physiology 23, 129--136.
doi:\href{https://doi.org/10.1093/treephys/23.2.129}{10.1093/treephys/23.2.129}

\hypertarget{ref-Niklas1994}{}
Niklas, K.J., 1994. Morphological evolution through complex domains of
fitness. Proceedings of the National Academy of Sciences 91, 6772--6779.
doi:\href{https://doi.org/10.1073/pnas.91.15.6772}{10.1073/pnas.91.15.6772}

\hypertarget{ref-Olesen2002a}{}
Olesen, J.E., Bindi, M., 2002. Consequences of climate change for
european agricultural productivity, land use and policy. European
journal of agronomy 16, 239--262.
doi:\href{https://doi.org/10.1016/s1161-0301(02)00004-7}{10.1016/s1161-0301(02)00004-7}

\hypertarget{ref-Paleari2015}{}
Paleari, L., Cappelli, G., Bregaglio, S., Acutis, M., Donatelli, M.,
Sacchi, G., Lupotto, E., Boschetti, M., Manfron, G., Confalonieri, R.,
2015. District specific, in silico evaluation of rice ideotypes improved
for resistance/tolerance traits to biotic and abiotic stressors under
climate change scenarios. Climatic Change 1--15.
doi:\href{https://doi.org/10.1007/s10584-015-1457-4}{10.1007/s10584-015-1457-4}

\hypertarget{ref-Pearcy2005}{}
Pearcy, R.W., Muraoka, H., Valladares, F., 2005. Crown architecture in
sun and shade environments: Assessing function and trade-offs with a
three-dimensional simulation model. New phytologist 166, 791--800.
doi:\href{https://doi.org/10.1111/j.1469-8137.2005.01328.x}{10.1111/j.1469-8137.2005.01328.x}

\hypertarget{ref-picheny2014multiobjective}{}
Picheny, V., 2014. Multiobjective optimization using gaussian process
emulators via stepwise uncertainty reduction. Statistics and Computing
on press.
doi:\href{https://doi.org/10.1007/s11222-014-9477-x}{10.1007/s11222-014-9477-x}

\hypertarget{ref-Picheny2017}{}
Picheny, V., Trépos, R., Casadebaig, P., 2017. Optimization of black-box
models with uncertain climatic inputs - application to sunflower
ideotype design. PloS One (in press).
doi:\href{https://doi.org/10.1371/journal.pone.0176815}{10.1371/journal.pone.0176815}

\hypertarget{ref-Poorter2006}{}
Poorter, H., Pepin, S., Rijkers, T., De Jong, Y., Evans, J.R., Körner,
C., 2006. Construction costs, chemical composition and payback time of
high-and low-irradiance leaves. Journal of Experimental Botany 57,
355--371.
doi:\href{https://doi.org/10.1093/jxb/erj002}{10.1093/jxb/erj002}

\hypertarget{ref-Poorter2006a}{}
Poorter, L., Bongers, L., Bongers, F., 2006. Architecture of 54
moist-forest tree species: Traits, trade-offs, and functional groups.
Ecology 87, 1289--1301.
doi:\href{https://doi.org/10.1890/0012-9658(2006)87\%5B1289:aomtst\%5D2.0.co;2}{10.1890/0012-9658(2006)87{[}1289:aomtst{]}2.0.co;2}

\hypertarget{ref-Quesnel2009}{}
Quesnel, G., Duboz, R., Ramat, É., 2009. The Virtual Laboratory
Environment -- An operational framework for multi-modelling, simulation
and analysis of complex dynamical systems. Simulation Modelling Practice
and Theory 17, 641--653.
doi:\href{https://doi.org/10.1016/j.simpat.2008.11.003}{10.1016/j.simpat.2008.11.003}

\hypertarget{ref-Quilot2016}{}
Quilot-Turion, B., Génard, M., Valsesia, P., Memmah, M.-M., 2016.
Optimization of allelic combinations controlling parameters of a peach
quality model. Frontiers in Plant Science 7, 1873.
doi:\href{https://doi.org/10.3389/fpls.2016.01873}{10.3389/fpls.2016.01873}

\hypertarget{ref-quilot.12}{}
Quilot-Turion, B., Ould-Sidi, M.-M., Kadrani, A., Hilgert, N., Génard,
M., Lescourret, F., 2012. Optimization of parameters of the ``virtual
fruit'' model to design peach genotype for sustainable production
systems. European Journal of Agronomy 42, 34--48.
doi:\href{https://doi.org/10.1016/j.eja.2011.11.008}{10.1016/j.eja.2011.11.008}

\hypertarget{ref-R2016}{}
R Core Team, 2016. R: A language and environment for statistical
computing. R Foundation for Statistical Computing, Vienna, Austria.

\hypertarget{ref-raquel.GECCO05}{}
Raquel, C.R., Naval, P.C., Jr., 2005. An effective use of crowding
distance in multiobjective particle swarm optimization, in: Proceedings
of the 7th Annual Conference on Genetic and Evolutionary Computation,
GECCO '05. ACM, New York, NY, USA, pp. 257--264.
doi:\href{https://doi.org/10.1145/1068009.1068047}{10.1145/1068009.1068047}

\hypertarget{ref-rasmussen2006gaussian}{}
Rasmussen, C., Williams, C., 2006. Gaussian processes for machine
learning. MIT Press.
doi:\href{https://doi.org/10.1007/978-3-540-28650-9_4}{10.1007/978-3-540-28650-9\_4}

\hypertarget{ref-Ray2012}{}
Ray, D.K., Ramankutty, N., Mueller, N.D., West, P.C., Foley, J.A., 2012.
Recent patterns of crop yield growth and stagnation. Nature
communications 3, 1293.
doi:\href{https://doi.org/10.1038/ncomms2296}{10.1038/ncomms2296}

\hypertarget{ref-renton2006similarities}{}
Renton, M., Guédon, Y., Godin, C., Costes, E., 2006. Similarities and
gradients in growth unit branching patterns during ontogeny in
`fuji'apple trees: A stochastic approach. Journal of Experimental Botany
57, 3131--3143.
doi:\href{https://doi.org/10.1093/jxb/erl075}{10.1093/jxb/erl075}

\hypertarget{ref-Ritchie1985}{}
Ritchie, J., Otter, S., 1985. Description and performance of
CERES-Wheat: A user-oriented wheat yield model. ARS-United States
Department of Agriculture, Agricultural Research Service (USA).

\hypertarget{ref-roustant2012dicekriging}{}
Roustant, O., Ginsbourger, D., Deville, Y., 2012. DiceKriging,
DiceOptim: Two R packages for the analysis of computer experiments by
kriging-based metamodeling and optimization. Journal of Statistical
Software 51, 1--55.
doi:\href{https://doi.org/10.18637/jss.v051.i01}{10.18637/jss.v051.i01}

\hypertarget{ref-Sadras1993}{}
Sadras, V.O., Villalobos, F.J., Fereres, E., Wolfe, D.W., 1993. Leaf
responses to soil water deficits: Comparative sensitivity of leaf
expansion and leaf conductance in field-grown sunflower. Plant and Soil
153, 189--194.

\hypertarget{ref-segura2008dissecting}{}
Segura, V., Cilas, C., Costes, E., 2008. Dissecting apple tree
architecture into genetic, ontogenetic and environmental effects. mixed
linear modelling of repeated spatial and temporal measures. New
Phytologist 178.
doi:\href{https://doi.org/10.1111/j.1469-8137.2007.02374.x}{10.1111/j.1469-8137.2007.02374.x}

\hypertarget{ref-semenov.14}{}
Semenov, M., Stratonovitch, P., Alghabari, F., Gooding, M., 2014.
Adapting wheat in europe for climate change. Journal of Cereal Science
59, 245--256.
doi:\href{https://doi.org/10.1016/j.jcs.2014.01.006}{10.1016/j.jcs.2014.01.006}

\hypertarget{ref-semenov.13}{}
Semenov, M.A., Stratonovitch, P., 2013. Designing high-yielding wheat
ideotypes for a changing climate. Food and Energy Security 2, 185--196.
doi:\href{https://doi.org/10.1002/fes3.34}{10.1002/fes3.34}

\hypertarget{ref-Sinclair2004}{}
Sinclair, T.R., Purcell, L.C., Sneller, C.H., 2004. Crop transformation
and the challenge to increase yield potential. Trends in Plant Science
9, 70--75.
doi:\href{https://doi.org/10.1016/j.tplants.2003.12.008}{10.1016/j.tplants.2003.12.008}

\hypertarget{ref-Srinivasan2017}{}
Srinivasan, V., Kumar, P., Long, S.P., 2017. Decreasing, not increasing,
leaf area will raise crop yields under global atmospheric change. Global
change biology 23, 1626--1635.
doi:\href{https://doi.org/10.1111/gcb.13526}{10.1111/gcb.13526}

\hypertarget{ref-Stockle2003}{}
Stockle, C.O., Donatelli, M., Nelson, R., 2003. CropSyst, a cropping
systems simulation model. European Journal of Agronomy 18, 289--307.
doi:\href{https://doi.org/10.1016/s1161-0301(02)00109-0}{10.1016/s1161-0301(02)00109-0}

\hypertarget{ref-sutton1996changing}{}
Sutton, T.B., 1996. Changing options for the control of deciduous fruit
tree diseases. Annual review of phytopathology 34, 527--547.

\hypertarget{ref-taylor2005biomechanics}{}
Taylor-Hell, J., 2005. Biomechanics in botanical trees
(Master's thesis). University of Calgary.

\hypertarget{ref-Technow2015}{}
Technow, F., Messina, C.D., Totir, L.R., Cooper, M., 2015. Integrating
crop growth models with whole genome prediction through approximate
bayesian computation. PloS one 10(6): e0130855.
doi:\href{https://doi.org/doi:10.1371/journal.pone.0130855}{doi:10.1371/journal.pone.0130855}

\hypertarget{ref-Tuck2006}{}
Tuck, G., Glendining, M.J., Smith, P., House, J.I., Wattenbach, M.,
2006. The potential distribution of bioenergy crops in europe under
present and future climate. Biomass and Bioenergy 30, 183--197.
doi:\href{https://doi.org/10.1016/j.biombioe.2005.11.019}{10.1016/j.biombioe.2005.11.019}

\hypertarget{ref-Valladares2007}{}
Valladares, F., Gianoli, E., Gómez, J.M., 2007. Ecological limits to
plant phenotypic plasticity. New Phytologist 176, 749--763.
doi:\href{https://doi.org/10.1111/j.1469-8137.2007.02275.x}{10.1111/j.1469-8137.2007.02275.x}

\hypertarget{ref-VanIttersum1997}{}
Van Ittersum, M., Rabbinge, R., 1997. Concepts in production ecology for
analysis and quantification of agricultural input-output combinations.
Field Crops Research 52, 197--208.
doi:\href{https://doi.org/10.1016/s0378-4290(97)00037-3}{10.1016/s0378-4290(97)00037-3}

\hypertarget{ref-Vanloqueren2009}{}
Vanloqueren, G., Baret, P.V., 2009. How agricultural research systems
shape a technological regime that develops genetic engineering but locks
out agroecological innovations. Research policy 38, 971--983.
doi:\href{https://doi.org/10.1016/j.respol.2009.02.008}{10.1016/j.respol.2009.02.008}

\hypertarget{ref-Violle2007}{}
Violle, C., Navas, M.-L., Vile, D., Kazakou, E., Fortunel, C., Hummel,
I., Garnier, E., 2007. Let the concept of trait be functional! Oikos
116, 882--892.
doi:\href{https://doi.org/10.1111/j.2007.0030-1299.15559.x}{10.1111/j.2007.0030-1299.15559.x}

\hypertarget{ref-White2003}{}
White, J.W., Hoogenboom, G., 2003. Gene-based approaches to crop
simulation. Agronomy Journal 95, 52--64.
doi:\href{https://doi.org/10.2134/agronj2003.0052}{10.2134/agronj2003.0052}

\hypertarget{ref-wu.12}{}
Wu, L., Le Dimet, F., De Reffye, P., Hu, B., Cournède, P., Kang, M.,
2012. An optimal control methodology for plant growth - case study of a
water supply problem of sunflower. Mathematics and computers in
simulation 82, 909--923.
doi:\href{https://doi.org/10.1016/j.matcom.2011.12.007}{10.1016/j.matcom.2011.12.007}

\hypertarget{ref-Yin2004}{}
Yin, X., Struik, P.C., Kropff, M.J., 2004. Role of crop physiology in
predicting gene-to-phenotype relationships. Trends in Plant Science 9,
426--432.
doi:\href{https://doi.org/10.1016/j.tplants.2004.07.007}{10.1016/j.tplants.2004.07.007}

\end{document}